\documentclass[12pt]{article}
\usepackage[dvips]{graphicx}
\usepackage{epsfig}
\usepackage{latexsym,amsmath,amsfonts,amssymb}
\usepackage[latin1]{inputenc}
\usepackage[american]{babel}
\usepackage[dvips]{graphicx}
\usepackage{bbm}
\def\np{Nucl. Phys.}
\def\pl{Phys. Lett.}
\def\prl{Phys. Rev. Lett.}
\def\pr{Phys. Rev.}

\def\im{Invent. Math.}

\def\jhep{J. High Energy Phys.}

\newcommand{\be}{\begin{equation}}
\newcommand{\ee}{\end{equation}}
\newcommand{\beq}{\begin{equation}}
\newcommand{\eeq}{\end{equation}}
\newcommand{\bea}{\begin{eqnarray}}
\newcommand{\eea}{\end{eqnarray}}

\newcommand{\ba}{\begin{eqnarray}}
\newcommand{\ea}{\end{eqnarray}}
\begin{document}
\baselineskip=15.5pt
\pagestyle{plain}
\setcounter{page}{1}
%--------+---------+---------+---------+---------+---------+---------+
%Body

% Ofer's definitions

\def\del{{\partial}}
\def\vev#1{\left\langle #1 \right\rangle}
\def\cn{{\cal N}}
\def\co{{\cal O}}
%\newfont{\Bbb}{msbm10 scaled 1200}     %instead of eusb10
%\newcommand{\mathbb}[1]{\mbox{\Bbb #1}}
\def\IC{{\mathbb C}}
\def\IR{{\mathbb R}}
\def\IZ{{\mathbb Z}}
\def\RP{{\bf RP}}
\def\CP{{\bf CP}}
\def\Poincare{{Poincar\'e }}
\def\tr{{\rm tr}}
\def\tp{{\tilde \Phi}}

\def\TL{\hfil$\displaystyle{##}$}
\def\TR{$\displaystyle{{}##}$\hfil}
\def\TC{\hfil$\displaystyle{##}$\hfil}
\def\TT{\hbox{##}}
\def\HLINE{\noalign{\vskip1\jot}\hline\noalign{\vskip1\jot}}
\def\seqalign#1#2{\vcenter{\openup1\jot
   \halign{\strut #1\cr #2 \cr}}}
\def\lbldef#1#2{\expandafter\gdef\csname #1\endcsname {#2}}
\def\eqn#1#2{\lbldef{#1}{(\ref{#1})}%
\begin{equation} #2 \label{#1} \end{equation}}
\def\eqalign#1{\vcenter{\openup1\jot
     \halign{\strut\span\TL & \span\TR\cr #1 \cr
    }}}
\def\eno#1{(\ref{#1})}
\def\href#1#2{#2}
\def\half{{1 \over 2}}

%--------+---------+---------+---------+---------+---------+---------+
%Hirosi's macros:
\def\ads{{\it AdS}}
\def\adsp{{\it AdS}$_{p+2}$}
\def\cft{{\it CFT}}

\newcommand{\ber}{\begin{eqnarray}}
\newcommand{\eer}{\end{eqnarray}}

\newcommand{\beqar}{\begin{eqnarray}}
\newcommand{\cN}{{\cal N}}
\newcommand{\cO}{{\cal O}}
\newcommand{\cA}{{\cal A}}
\newcommand{\cT}{{\cal T}}
\newcommand{\cF}{{\cal F}}
\newcommand{\cC}{{\cal C}}
\newcommand{\cR}{{\cal R}}
\newcommand{\cW}{{\cal W}}
\newcommand{\eeqar}{\end{eqnarray}}
\newcommand{\tht}{\thteta}
\newcommand{\lm}{\lambda}\newcommand{\Lm}{\Lambda}
\newcommand{\eps}{\epsilon}

%--------+---------+---------+---------+---------+---------+---------+

\newcommand{\nonu}{\nonumber}
\newcommand{\oh}{\displaystyle{\frac{1}{2}}}
\newcommand{\dsl}
   {\kern.06em\hbox{\raise.15ex\hbox{$/$}\kern-.56em\hbox{$\partial$}}}
\newcommand{\id}{i\!\!\not\!\partial}
\newcommand{\as}{\not\!\! A}
\newcommand{\ps}{\not\! p}
\newcommand{\ks}{\not\! k}
\newcommand{\D}{{\cal{D}}}
\newcommand{\dv}{d^2x}
\newcommand{\Z}{{\cal Z}}
\newcommand{\N}{{\cal N}}
\newcommand{\Dsl}{\not\!\! D}
\newcommand{\Bsl}{\not\!\! B}
\newcommand{\Psl}{\not\!\! P}
\newcommand{\eeqarr}{\end{eqnarray}}
\newcommand{\ZZ}{{\rm \kern 0.275em Z \kern -0.92em Z}\;}
%--------------------------------Alfonso's definitions%%%%%%%%%%%%%

% DEFINITIONS

\def\del{{\delta^{\hbox{\sevenrm B}}}} \def\ex{{\hbox{\rm e}}}
\def\azb{A_{\bar z}} \def\az{A_z} \def\bzb{B_{\bar z}} \def\bz{B_z}
\def\czb{C_{\bar z}} \def\cz{C_z} \def\dzb{D_{\bar z}} \def\dz{D_z}
\def\im{{\hbox{\rm Im}}} \def\mod{{\hbox{\rm mod}}} \def\tr{{\hbox{\rm Tr}}}
\def\ch{{\hbox{\rm ch}}} \def\imp{{\hbox{\sevenrm Im}}}
\def\trp{{\hbox{\sevenrm Tr}}} \def\vol{{\hbox{\rm Vol}}}
\def\rl{\Lambda_{\hbox{\sevenrm R}}} \def\wl{\Lambda_{\hbox{\sevenrm W}}}
\def\fc{{\cal F}_{k+\cox}} \def\vev{vacuum expectation value}
\def\nodiv{\mid{\hbox{\hskip-7.8pt/}}}
\def\ie{{\em i.e.}}
\def\ie{\hbox{\it i.e.}}

\def\CC{{\mathchoice
{\rm C\mkern-8mu\vrule height1.45ex depth-.05ex
width.05em\mkern9mu\kern-.05em}
{\rm C\mkern-8mu\vrule height1.45ex depth-.05ex
width.05em\mkern9mu\kern-.05em}
{\rm C\mkern-8mu\vrule height1ex depth-.07ex
width.035em\mkern9mu\kern-.035em}
{\rm C\mkern-8mu\vrule height.65ex depth-.1ex
width.025em\mkern8mu\kern-.025em}}}

\def\RR{{\rm I\kern-1.6pt {\rm R}}}
\def\NN{{\rm I\!N}}
\def\ZZ{{\rm Z}\kern-3.8pt {\rm Z} \kern2pt}
\def\IB{\relax{\rm I\kern-.18em B}}
\def\ID{\relax{\rm I\kern-.18em D}}
\def\II{\relax{\rm I\kern-.18em I}}
\def\IP{\relax{\rm I\kern-.18em P}}
\newcommand{\CS}{{\scriptstyle {\rm CS}}}
\newcommand{\CSs}{{\scriptscriptstyle {\rm CS}}}
\newcommand{\rc}{\nonumber\\}
\newcommand{\bear}{\begin{eqnarray}}
\newcommand{\eear}{\end{eqnarray}}
\newcommand{\W}{{\cal W}}
\newcommand{\F}{{\cal F}}
\newcommand{\x}{{\cal O}}
\newcommand{\LL}{{\cal L}}

\def\mani{{\cal M}}
\def\calo{{\cal O}}
\def\calb{{\cal B}}
\def\calw{{\cal W}}
\def\calz{{\cal Z}}
\def\cald{{\cal D}}
\def\calc{{\cal C}}
\def\to{\rightarrow}
\def\ele{{\hbox{\sevenrm L}}}
\def\ere{{\hbox{\sevenrm R}}}
\def\zb{{\bar z}}
\def\wb{{\bar w}}
\def\nodiv{\mid{\hbox{\hskip-7.8pt/}}}
\def\menos{\hbox{\hskip-2.9pt}}
\def\dr{\dot R_}
\def\drr{\dot r_}
\def\ds{\dot s_}
\def\da{\dot A_}
\def\dga{\dot \gamma_}
\def\ga{\gamma_}
\def\dal{\dot\alpha_}
\def\al{\alpha_}
\def\cl{{closed}}
\def\cls{{closing}}
\def\vev{vacuum expectation value}
\def\tr{{\rm Tr}}
\def\to{\rightarrow}
\def\too{\longrightarrow}

% Umut likes:

\def\a{\alpha}
\def\b{\beta}
\def\c{\gamma}
\def\d{\delta}
\def\e{\epsilon}           % Also, \varepsilon
\def\f{\phi}               %      \varphi
\def\vf{\varphi}  \def\tvf{\tilde{\varphi}}
\def\vp{\varphi}
\def\g{\gamma}
\def\h{\eta}
\def\i{\iota}
\def\j{\psi}
\def\k{\kappa}                    % Also, \varkappa (see below)
\def\l{\lambda}
\def\m{\mu}
\def\n{\nu}
\def\o{\omega}  \def\w{\omega}
\def\q{\theta}  \def\th{\theta}                  %     \vartheta
\def\r{\rho}                                     %     \varrho
\def\s{\sigma}                                   %     \varsigma
\def\t{\tau}
\def\u{\upsilon}
\def\x{\xi}
\def\z{\zeta}
\def\pt{\tilde{\varphi}}
\def\tt{\tilde{\theta}}
\def\lab{\label}
\def\6{\partial}
\def\wg{\wedge}
\def\atanh{{\rm arctanh}}
\def\bpsi{\bar{\psi}}
\def\bt{\bar{\theta}}
\def\bvf{\bar{\varphi}}

%
% FONTS

%\newfont{\headfont}{cmbx10 scaled 1440}
\newfont{\namefont}{cmr10}
%\newfont{\initialfont}{cmr10 scaled 1200}
\newfont{\addfont}{cmti7 scaled 1440}
\newfont{\boldmathfont}{cmbx10}
%\newfont{\figfont}{cmr7 scaled 1200}
\newfont{\headfontb}{cmbx10 scaled 1728}
%%%%%%%%%%%%%%%%%%%%%%%%%%%%%%%%%%%%%%%%%%%%%%%%%%%%%%%%%%%%%%%%%%%%%%%%%%%%
%%%%%%%%%%%%%%%Stefano and Francesco fonts%%%%%%%%%%%%%%%%%%%%%%%%%%%%%%%%
%%%%%%%%%%%%%%%%%%%%%%%%%%%%%%%%%%%%%%%%%%%%%%%%%%%%%%%%%%%%%%%%%%%%%%%%%%%%
\newcommand{\re}{\,\mathbb{R}\mbox{e}\,}
\newcommand{\hyph}[1]{$#1$\nobreakdash-\hspace{0pt}}
\providecommand{\abs}[1]{\lvert#1\rvert}
\newcommand{\Nugual}[1]{$\mathcal{N}= #1 $}
\newcommand{\sub}[2]{#1_\text{#2}}
\newcommand{\partfrac}[2]{\frac{\partial #1}{\partial #2}}
\newcommand{\bsp}[1]{\begin{equation} \begin{split} #1 \end{split} \end{equation}}
\newcommand{\calF}{\mathcal{F}}
\newcommand{\calO}{\mathcal{O}}
\newcommand{\calM}{\mathcal{M}}
\newcommand{\calV}{\mathcal{V}}
\newcommand{\bbZ}{\mathbb{Z}}
\newcommand{\bbC}{\mathbb{C}}

\numberwithin{equation}{section}

\newcommand{\Tr}{\mbox{Tr}}    % trace over gauge indices

%%%%%%%%%%%%%%%%%%%%%%%%%%%%%%%%%%%%%%%%%%%%%%%%%%%%%%%%%%%%%%%%%%%%%%%%%%%%
%%%%%%%%%%%%%%%%%%%%%%%%%%%%%%%%%%%%%%%%%%%%%%%%%%%%%%%%%%%%%%%%%%%%%%%%%%%%

%
\renewcommand{\theequation}{{\rm\thesection.\arabic{equation}}}
\begin{titlepage}
%
%\rightline{US-FT-3/07}
\vspace{0.1in}

\begin{center}
\Large \bf The supergravity dual of 3d supersymmetric  gauge theories with unquenched flavors
\end{center}
\vskip 0.2truein
\begin{center}
Felipe Canoura\footnote{canoura@fpaxp1.usc.es}, Paolo Merlatti
\footnote{merlatti@fpaxp1.usc.es}
 and
Alfonso V. Ramallo\footnote{alfonso@fpaxp1.usc.es}\\
\vspace{0.2in}
\it{
Departamento de  Fisica de Particulas, Universidade
de Santiago de
Compostela\\and\\Instituto Galego de Fisica de Altas
Enerxias (IGFAE)\\E-15782, Santiago de Compostela, Spain
}
\vspace{0.2in}
\end{center}
\vspace{0.2in}
%\begin{center}
\centerline{{\bf Abstract}}
We obtain the supergravity dual of ${\cal N}=1$ gauge theory in 2+1 dimensions with a large number of unquenched massless flavors. The geometries found are obtained by solving the equations of motion of supergravity coupled to a suitable continuous distribution of flavor  branes. The background obtained preserves two supersymmetries. We find that when $N_c\ge 2N_f$ the behavior of the solutions is compatible with having an asymptotically free dual gauge theory with dynamical quarks. On the contrary, when $N_c<2N_f$ the theory develops a Landau pole in the UV. We also find a new family of (unflavored) backgrounds generated by D5-branes that wrap a three-cycle of a cone with $G_2$ holonomy.

\smallskip
\end{titlepage}
\setcounter{footnote}{0}

\tableofcontents

%--------+---------+---------+---------+---------+---------+---------+
%Body

\newpage

\section{Introduction}
The gauge/gravity correspondence \cite{jm,MAGOO} has provided us with a very powerful tool to explore the dynamics of gauge theories at strong coupling. In its original formulation the correspondence is a duality between the $AdS_5\times S^5$ background of type IIB supergravity and ${\cal N}=4, d=4$ super Yang-Mills theory, in which all fields transform in the adjoint representation of the gauge group. Clearly, to extend this duality to systems closer to the particle physics phenomenology we should be able to add matter fields transforming in the fundamental representation of the gauge group. This is equivalent to adding open string degrees of freedom to the supergravity side of the correspondence.

It was originally proposed in \cite{KK} that such an open string sector can be obtained by adding certain D-branes to the supergravity background. If the number $N_f$ of such flavor branes is small compared with the number $N_c$ of colors, we can treat the flavor branes as probes in the background created by the color branes. This is the so-called quenched approximation, which corresponds, in the field theory side, to suppressing quark loops by factors $1/N_c$ in the 't Hooft large $N_c$ expansion. The fluctuation modes of the probe D-brane in the supergravity background provide a holographic description of the flavor sector of the gauge theory and one can extract the corresponding meson spectrum by analyzing the normalizable fluctuations of the probe \cite{KMMW} (for a review and a list of references, see \cite{Erdmenger:2007cm}). 

When the number $N_f$ of flavors is of the same order as the number $N_c$ of colors, the backreaction of the flavor branes on the metric can no longer be neglected. On the field theory side the inclusion of the backreaction is equivalent to considering the so-called Veneziano limit, in which $N_c$ and $N_f$ are large and their ratio $N_f/N_c$ is fixed. 
In this limit quark loops are no longer suppressed. In the last few years there have been several attempts to construct supergravity duals of these unquenched systems, both for four-dimensional \cite{4dlocalized} and three-dimensional gauge theories \cite{3dlocalized},  by using  solutions of supergravity generated by localized intersections of branes.

Recently, a different approach has been proposed in ref. \cite{Casero:2006pt}. Instead of solving the equations of pure supergravity, the authors of \cite{Casero:2006pt} considered the full gravity plus (flavor) branes system.  The action of such a system contains the Dirac-Born-Infeld action of the flavor branes, which governs their worldvolume dynamics and their coupling to the different supergravity fields. 
Notice that this is consistent with the fact that color branes undergo a geometric transition and are converted into fluxes, whereas, on the contrary, flavor branes are still present after the geometric transition. Thus, from a conceptual point of view, color and flavor branes are not  equivalent and, therefore, should be treated differently. By considering a suitable continuous distribution of flavor branes the authors of \cite{Casero:2006pt} were able to find a set of BPS equations and to solve them numerically (see \cite{noncritical} for a similar approach in the context of non-critical string theory). The resulting solution is the flavored backreacted version of the background found in \cite{CV} and proposed in \cite{MN} as the supergravity dual of ${\cal N}=1$ super Yang-Mills theory in four dimensions. Further developments of this approach can be found in refs. \cite{Casero:2007pz}-\cite{Caceres:2007mu}. In this paper we will apply this circle of ideas to the case of ${\cal N}=1$ gauge theories in 2+1 dimensions. The corresponding unflavored supergravity dual was found in ref. \cite{Chamseddine:2001hk}, and it was interpreted as being generated by D5-branes wrapped on a three-cycle of a manifold of $G_2$ holonomy in \cite{Maldacena:2001pb} (see also \cite{Schvellinger:2001ib}-\cite{Bertoldi:2002ks}).

The low number of supersymmetries preserved by the solution of \cite{Chamseddine:2001hk} (just two real supercharges) is a nice feature and makes it appealing also from the perspective of its dual field theory. As pointed out in ref. \cite{Maldacena:2001pb},
 this theory reduces in the IR to 2+1 dimensional   ${\cal N}=1$ supersymmetric $U(N_c)$Yang-Mills theory with a level $k$ Chern-Simons interaction. Such theory coupled to an adjoint massive scalar field should arise on the domain walls separating the different vacua of pure ${\cal N}=1$ super-Yang-Mills in 3+1 dimensions. For $k\neq 0$ the theory has a mass gap, at least classically, with mass of order $g_{YM}^2 |k|$. This implies that for $|k|>>1$, \ie\ when  we can trust the classical result,  there are no Goldstone fermions and, therefore, supersymmetry is unbroken. Actually, the Witten index for such a theory was computed in  \cite{Witten:1999ds}, where it has been shown  that for $k\geq  N_c/2$ supersymmetry is unbroken, while it is broken for $k<N_c/2$. In the borderline case ($k=N_c/2$)  there is just one supersymmetric vacuum. Being the supergravity solution of \cite{Chamseddine:2001hk} supersymmetric and without parameters that could label different vacua, it is reasonable to expect  that the  dual field theory is the one describing the $k=N_c/2$ case. It was shown  in ref. \cite{Maldacena:2001pb} that  this is actually the case.

We will start our analysis by generalizing the ansatz of \cite{Chamseddine:2001hk} for the unflavored solutions. This generalization will allow us to find a new class of solutions in which, in the UV,  the metric becomes asymptotically the direct product of  a $G_2$ cone  and a three-dimensional Minkowski space, while the dilaton becomes constant. This is  in contrast to the background of \cite{Chamseddine:2001hk}, in which the dilaton grows linearly with the holographic coordinate. For this generalized ansatz we will be able to find a system of first-order BPS equations which ensure that our solutions preserve two supersymmetries. We will perform a careful analysis of the regularity conditions to be imposed on the functions of our ansatz, which will allow us to fix some parameters of our solutions and to determine the appropriate initial conditions needed to solve the BPS differential equations. The new solutions, which are found numerically, can be naturally interpreted as non-near horizon versions of the one of \cite{Chamseddine:2001hk}.

After completing the analysis of the unflavored backgrounds, we will study the addition of flavor D5-branes. First of all, we will use kappa symmetry \cite{swedes} to determine a continuous family of embeddings of probes that preserve all the supersymmetries of the background and which can be used as flavor branes for massless quarks. These embeddings have the topology of a cylinder and are very similar to the ones found in \cite{Nunez:2003cf} (and used in \cite{Casero:2006pt}) in the case of the supergravity dual of ${\cal N}=1$ gauge theories in four dimensions. It turns out that the embeddings we will  find can be straightforwardly smeared in their transverse directions without breaking supersymmetry.  Moreover,  one can combine them in a way compatible with our  generalized metric ansatz. We will use this fact to compute the backreacted geometry. 

As the flavor branes act as a source of the RR forms in the backreacted solution, we will have  to modify the ansatz of the RR three-form to include the violation of its Bianchi identity in a very precise form.  After this modification of the ansatz, we will look again at the BPS equations that enforce supersymmetry and we will get a system of differential equations that generalizes the one found for the unflavored system. These equations depend now both on $N_c$ and $N_f$ and can be solved by imposing regularity conditions that are similar to the ones used for the unflavored case. By solving the BPS equations for different numbers of colors and  flavors we will discover that  the system behaves differently depending on whether $N_c$ is larger or smaller than $2N_f$. The most interesting case occurs when $N_c\ge 2N_f$. In this regime the behavior of the solution is compatible with having an asymptotically free gauge theory with dynamical massless quarks. We will  confirm this result by computing, from our solution,  the beta function and the quark-antiquark potential energy. We will get the expected linear confining potential and the dual description of the confining string breaking due to pair creation.  On the contrary, when $N_c<2N_f$ the solution ceases to exist beyond some value of the holographic coordinate. This behavior is compatible with having a Landau pole in the UV.

This paper is organized as follows. In section \ref{Unflavor-section} we will formulate our generalized ansatz for the unflavored case. The corresponding BPS equations are obtained in appendix \ref{BPSapp}. It turns out that these equations admit a truncation, in which some functions of the ansatz are fixed to some particular values and, as a consequence, the system of BPS equations greatly simplifies. Due to this simplification we will first analyze this truncated system in subsection \ref{truncated-unflavor-section}. In subsection \ref{untruncated-unflavored-subsection} we will consider the full system which, in general, presents a better IR behavior. In this subsection we carefully examine the regularity conditions to be imposed on the solutions of the BPS equations.

In section \ref{flavor-addition-section} we consider the addition of flavor branes. We first determine the kappa symmetric cylinder embeddings and then we find the particular distribution of them that preserves supersymmetry and is compatible with our metric ansatz. This distribution dictates the modification of the ansatz of the RR three-form needed to encompass the modification of the Bianchi identity induced by the flavor branes. The corresponding BPS equations  for this case are also found in appendix
\ref{BPSapp} while, in appendix \ref{EOMapp} we verify that, quite remarkably, the first-order equations derived from supersymmetry imply the fulfillment    of the second order equations of motion for the coupled gravity plus branes theory. It turns out that the BPS system with flavor admits the same truncation as in the $N_f=0$ case. We study this truncated system in section \ref{truncated-flavor}. The full system for $N_c\ge 2N_f$  is analyzed in section \ref{untruncated-flavor}, whereas section \ref{untruncated-flavor-Landau} is devoted to the study of this same system when $N_c<2N_f$.

Finally, in section \ref{conclusions} we recapitulate our results and discuss some possible extensions of our work.

\section{Deforming the unflavored solution}
\label{Unflavor-section}

Let us begin by describing in detail  the ansatz that we will adopt for the unflavored backgrounds we are interested in. As a particular case the family of our solutions will include the one found  originally in \cite{Chamseddine:2001hk} and interpreted in \cite{Maldacena:2001pb} as a supergravity dual of ${\cal N}=1$ super Yang Mills theory in 2+1 dimensions.  More concretely,  let $\sigma^i$ and $\omega^i$ $(i=1,2,3)$ be two sets of SU(2) left-invariant one forms, obeying:
\beq
d\sigma^i=-{1\over 2}\,\epsilon_{ijk}\,\sigma^j\wedge \sigma^k\,\,,
\qquad\qquad
d\omega^i=-{1\over 2}\,\epsilon_{ijk}\,\omega^j\wedge \omega^k\,\,.
\label{sigma-w}
\eeq
The forms $\sigma^i$ and $\omega^i$ parameterize two three-spheres. In the geometries we  will be dealing with,  these spheres are fibered by a one-form $A^i$. The corresponding  ten-dimensional metric of the type IIB theory in the Einstein frame is given by:
\beq
ds^2\,=\,e^{2f}\left[ dx_{1,2}^2+dr^2+\frac{e^{2h}}{4}(\sigma^i)^2+\frac{e^{2g}}{4}(\omega^i-A^i)^2\right]\,\,,
\label{ansatz}
\eeq
where $dx_{1,2}^2$ is the Minkowski metric in 2+1 dimensions, $r$ is a radial (holographic) coordinate and $f$, $g$ and $h$ are functions of  $r$.  In addition, the one-form $A^i$ will be taken as:
\beq
A^i\,=\,{1+w(r)\over 2}\,\,\sigma^i\,\,,
\label{Ai-w}
\eeq
with $w(r)$ being a new function of $r$. The backgrounds considered here are also endowed with a non-trivial dilaton $\phi$ and an RR three-form $F_{3}$, which we will take as:
\beq
\frac{F_{3}}{N_c}=-\frac{1}{4}(\omega^1-B^1)\wedge(\omega^2-B^2)\wedge(\omega^3-B^3)+\frac{1}{4}F^i\wedge(\omega^i-B^i)+H\,\,,
\label{F3ansatz}
\eeq
where $B^i$ is a new one-form and $F^i$ are the components of its field strength, given by:
\beq
F^i\,=\,dB^i\,+\,{1\over 2}\,\epsilon_{ijk}\,B^j\wedge B^k\,\,.
\label{Fi}
\eeq
In (\ref{F3ansatz}) $H$ is a three-form that is determined by imposing the Bianchi identity for $F_3$, namely:
\beq
dF_3\,=\,0\,\,.
\label{Bianchi-id}
\eeq
By using (\ref{sigma-w}) one can easily  check  from the explicit expression written in 
(\ref{F3ansatz}) that, in order to fulfill (\ref{Bianchi-id}), the three-form $H$ must satisfy the equation:
\beq
dH\,=\,{1\over 4}\,F^i\wedge F^i\,\,.
\label{dH}
\eeq
In what follows we shall adopt the following ansatz for $B^i$:
\beq
B^i\,=\,{1+\gamma(r)\over 2}\,\,\sigma^i\,\,,
\label{Bi}
\eeq
where $\gamma(r)$ is a new function.  After plugging the ansatz of $B^i$ written in (\ref{Bi}) into (\ref{Fi}), one gets the expression of $F^i$ in terms of $\gamma(r)$, \ie:
\beq
F^i\,=\,{\gamma'\over 2}\,\,dr\wedge \sigma^i\,+\,
{\gamma^2-1\over 8}\,\,\epsilon_{ijk}\,\sigma^j\wedge \sigma^k\,\,,
\label{Fi-gamma}
\eeq
where the prime denotes the derivative with respect to the radial variable $r$. Using this result for $F^i$ in (\ref{dH}) one can easily determine the three-form $H$ in terms of $\gamma$. Let us parameterize $H$  as:
\beq
H\,=\,{1\over 32}\,\,{1\over 3!}\,\,{\cal H}(r)\,\,\epsilon_{ijk}\,\,
\sigma^i\wedge \sigma^j\wedge\sigma^k\,\,.
\eeq 
Then, by solving (\ref{dH}) for $H$,  one can verify that 
${\cal H}(r)$ is the following function of the radial variable:
\beq
{\cal H}\,=\,2\gamma^3-6\gamma\,+\,8\kappa\,\,,
\label{calH}
\eeq
with $\kappa$ being an integration constant. 

In the particular case in which the function $g$ is constant and the fibering functions 
$w$ and $\gamma$ are equal our ansatz reduces to the one considered in refs. \cite{Chamseddine:2001hk,Maldacena:2001pb}. Actually, we will verify that the BPS equations fix, in this case,  the constant  value of $g$ to be $e^{2g}=N_c$. Moreover, this type of solution is naturally obtained by considering a fivebrane wrapped on a three-sphere in seven dimensional gauged supergravity. This three-sphere of the seven dimensional solution is just the one parameterized by the $\sigma^i$'s, while $A^i=B^i$ is the $SU(2)$ gauge field of the gauged supergravity and $H$ the corresponding three-form. The expressions (\ref{ansatz}) and (\ref{F3ansatz}) for the metric and RR three-form of our ansatz are just the ones that are obtained naturally upon uplifting the solution from seven to ten dimensions. Notice that $N_c$ characterizes the flux of the RR three-form and it corresponds to the number of colors on the gauge theory side, whereas the constant $\kappa$ of (\ref{calH})  is, in the analysis of \cite{Maldacena:2001pb}, related to the coefficient  of the Chern-Simons term in the 2+1 dimensional gauge theory.

By requiring that our background preserves some fraction of supersymmetry we arrive at a system of first-order BPS equations for the different functions of  our ansatz. In its full generality this analysis is rather involved and it is presented in detail in appendix \ref{BPSapp}. Let us mention here that the number of supersymmetries preserved by our solutions is equal to two, which is the right amount of SUSY expected for an ${\cal N}=1$ gauge theory in 2+1 dimensions. Moreover, supersymmetry imposes the following relation between the dilaton $\phi$ and the function $f$ appearing in the metric (\ref{ansatz}):
\beq
\phi\,=\,4f\,\,.
\eeq

\subsection{The truncated system}
\label{truncated-unflavor-section}

As mentioned above the equations imposed by supersymmetry on the functions of our ansatz are obtained in appendix \ref{BPSapp}. By inspecting these equations one can check that  they admit solutions in which the fibering functions $w$ and $\gamma$ vanish, as well as the integration constant $\kappa$, namely:
\beq
w=\gamma=0\,\,,\qquad\qquad
\kappa=0\,\,.
\label{truncation}
\eeq
By performing the truncation (\ref{truncation}) the first-order BPS system simplifies drastically. Actually, one can verify that it reduces to the following three equations for $\phi=4f$, $h$ and $g$:
\bear
&&\phi'\,=\,N_c\,e^{-g}\,\Big[\,e^{-2g}\,-\,{3\over 4}\,e^{-2h}\,\Big]\,\,,\rc\rc
&&h'\,=\,{e^{-2h}\over 2}\,\Big[\,e^g\,+\,N_c\,e^{-g}\,\Big]\,\,,\rc\rc
&&g'\,=\,\Big[\,{e^{-2h}\over 4}\,-\,e^{-2g}\,\Big]\,\Big[\,N_c\,e^{-g}\,-\,e^{g}\,\Big]\,\,.
\,\,
\label{ab-system}
\eear
To integrate this system, let us consider first the possibility of having  solutions with $g$ constant. It follows from the equation for $g'$ written in (\ref{ab-system}) that $g$ must be such that:
\beq
e^{2g}\,=\,N_c\,\,.
\label{e2g=Nc}
\eeq
Plugging this value of $g$ in the second equation in (\ref{ab-system}) one easily shows that  the equation for $h$ becomes:
\beq
h'\,=\,\sqrt{N_c}\,e^{-2h}\,\,,
\eeq
which can be integrated immediately as:
\beq
e^{2h}\,=\,2\sqrt{N_c}\,r\,\,.
\label{h-for-e2g=Nc}
\eeq
Using these values of $g$ and $h$ the dilaton can be readily obtained from the first equation in (\ref{ab-system}), namely:
\beq
e^{2\phi}\,=\,e^{2\phi_0}\,{e^{{2r\over \sqrt{N_c}}}\over r^{{3\over 4}}}\,\,,
\label{phi-for-e2g=Nc}
\eeq
where $\phi_0$ is a constant. The solution given by eqs.  (\ref{e2g=Nc}), (\ref{h-for-e2g=Nc}) and (\ref{phi-for-e2g=Nc}) was obtained in \cite{Acharya:2000mu} as the background generated by fivebranes wrapped on a three-cycle of a manifold of $G_2$ holonomy.  The corresponding metric is singular at $r=0$. Notice also that the dilaton (\ref{phi-for-e2g=Nc}) grows linearly with the holographic coordinate for $r\to\infty$, as it should for  a background created by fivebranes in the near-horizon limit.

In order to study the system (\ref{ab-system}) in general and find other classes of solutions, let us define  a new radial variable $\rho$ as:
\beq
\rho\equiv e^{2h}\,\,,
\label{rho}
\eeq
and a new function $F$ as:
\beq
F\,\equiv\,e^{2g}\,\,.
\label{F}
\eeq
We will consider $F$ as a function of $\rho$. From the equations for $h'$ and $g'$ written in (\ref{ab-system}) we get the following equation for $F(\rho)$:
\beq
{dF\over d\rho}\,=\,\Big[\,2\,-\,{F\over 2\rho}\,\Big]\,
{F-N_c\over F+N_c}\,\,,
\label{F-rho}
\eeq
while the equation for the dilaton is:
\beq
{d\phi\over d\rho}\,=\,{N_c\,\over F(F+N_c)}\,\,
\Big[\,1\,-\,{3F\over 4\rho}\,\Big]\,\,.
\label{dilaton-rho}
\eeq
From the equation for $h$ in the system (\ref{ab-system}) it is straightforward to verify that the jacobian of the change of radial variable is:
\beq
{dr\over d\rho}\,=\,{\sqrt{F}\over F+N_c}\,\,,
\label{r-rho-unflavored}
\eeq
and, thus,  one can write the metric as:
\beq
ds^2\,=\,e^{{\phi\over 2}}\,\Big[\,dx^2_{1,2}\,+\,{F\over (F+N_c)^2}\,(d\rho)^2\,+\,
{\rho\over 4}\,(\sigma^i)^2\,+\,{F\over 4}\,\Big(\,\omega^i\,-\,{\sigma^i\over 2}\,\Big)^2
\,\Big]\,\,.
\label{metric-in-rho-ab}
\eeq
By inspecting (\ref{F-rho}) we recognize our special solution (\ref{e2g=Nc})-(\ref{phi-for-e2g=Nc}) as the one that is obtained by taking $F=N_c$ in (\ref{F-rho}) and (\ref{dilaton-rho}). Another case in which the BPS equations can be solved analytically is when $N_c=0$. Indeed, in this case the RR three-form vanishes, the dilaton is constant and
(\ref{F-rho}) becomes:
\beq
{dF\over d\rho}\,+\,{F\over 2\rho}\,=\,2\,\,,
\qquad\qquad (N_c=0)\,\,.
\label{FNc=0}
\eeq
The general solution of (\ref{FNc=0}) can be found easily:
\beq
F\,=\,{4\over 3}\,\,\rho\,+\,{c\over \rho^{{1\over 2}}}\,\,,
\qquad\qquad (N_c=0)\,\,,
\label{FNc=0-sol}
\eeq
where $c$ is an integration constant. Let us write the form of this solution in a more suitable form. For this purpose it is convenient to perform a new change in the radial variable, namely:
\beq
\rho\,=\,{1\over 3}\,\tau^2\,\,,
\eeq
and to define the constant $a$, related to the integration constant $c$  in (\ref{FNc=0-sol}) as  $c=-4a^3/9\sqrt{3}$.  In terms of these  quantities the  metric  (\ref{metric-in-rho-ab}) becomes:
\beq
ds^2\,=\,e^{{\phi_{*}\over 2}}\,\Big[\,dx^2_{1,2}\,+\,
{(d\tau)^2\over 1-{a^3\over \tau^3}}
\,+\,
{\tau^2\over 12}\,(\sigma^i)^2\,+\,{\tau^2\over 9}\,\,
\Big(\,1-{a^3\over \tau^3}\,\Big)\,
\Big(\,\omega^i\,-\,{\sigma^i\over 2}\,\Big)^2
\,\Big]\,\,,
\label{UV-asymp-metric}
\eeq
where $\phi_*$ is the constant value of the dilaton. Notice that 
(\ref{UV-asymp-metric}) is just the metric of the direct product of a 2+1 Minkowski space and a  manifold of $G_2$ holonomy. This metric of $G_2$ holonomy is just the well-known Bryant-Salamon metric \cite{bs}, which has the topology of $\RR^4\times S^3$ and  asymptotes  to a $G_2$ cone for large values of the radial coordinate $\tau$. Notice that  $\tau\ge a$ in (\ref{UV-asymp-metric}) and as $\tau\to a$ one of the two three-spheres shrinks to a point while the other remains finite.  When $a=0$ this manifold is singular at the origin $\tau=0$. This singularity is cured by switching on a non-zero value of the parameter $a$, in a way very similar to that which happens to the resolved conifold. 

\begin{figure}[ht]
\begin{center}
\includegraphics[width=0.95\textwidth]{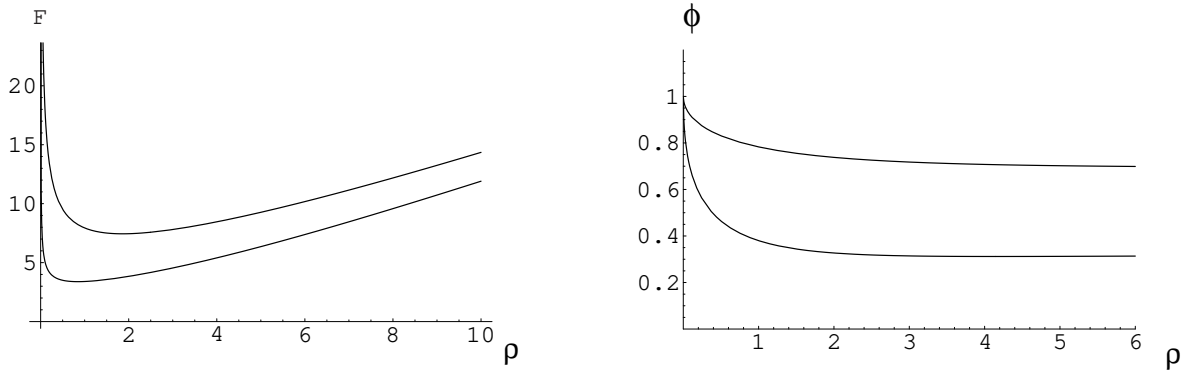}
\end{center}
\caption[UnF-Phi.eps]{$F$ and $\phi$ for two different values of the constant $c_0$  of eqs. (\ref{F-vs-rho-IR}) and (\ref{Phi-vs-rho-IR}) ($c_0= 5 $ and $10$) and $N_c=1$. } 
\label{UnF-Phi}
\end{figure}

Having obtained the previous solutions for near-horizon fivebranes and (resolved) $G_2$ cones without branes,  it is quite natural to look at solutions with RR three-form whose metric becomes in the UV the direct product of 2+1 Minkowski space and a $G_2$ cone.   In a sense these solutions  would correspond to going beyond the near-horizon region of the fivebrane background.  In terms of the variables $F$ and $\rho$ it is clear that we are looking for solutions such that:
\beq
F\sim {4\over 3}\,\rho\,+\,\cdots\,\,,\qquad\qquad
(\rho\to\infty)\,\,.
\label{F-UVexplicit}
\eeq
Notice that, when (\ref{F-UVexplicit}) holds, $1-3F/ 4\rho\approx 0$ for 
$\rho\to\infty$ and, therefore, 
eq. (\ref{dilaton-rho}) shows that 
the dilaton is stabilized in the UV, \ie\ 
$\phi\,\to\,{\rm constant}$ for large $\rho$, in contrast to what happens in (\ref{phi-for-e2g=Nc}). Actually, one can show that for large $\rho$ the solution of the differential equation (\ref{F-rho}) that behaves as in (\ref{F-UVexplicit}) can be approximated as:
\beq
F\,=\,{4\over 3}\,\rho\,-\,4N_c\,+\,{15 N_c^2\over \rho}\,-\,{45N_c^3\over 4}\,{1\over \rho^2}\,+\,\cdots\,\,,
\qquad\qquad (\rho\to\infty)\,\,.
\label{G2-F-unflavored}
\eeq
By plugging this expansion in the equation (\ref{dilaton-rho}) for the dilaton, one gets the  following UV expansion:
\beq
{d\phi\over d\rho}\,=\,{27 N_c^2\over 16}\,\,{1\over \rho^3}\,+\,{81 N_c^3\over 32}\,
{1\over \rho^4}\,+\,\cdots
\,\,,\qquad\qquad (\rho\to\infty)
\,\,,
\label{G2-phi-unflavored}
\eeq
which can be integrated as:
\beq
\phi\,=\,\phi_{\infty}\,-\,{27 N_c^2\over 32}\,\,{1\over \rho^2}\,+\,\cdots
\,\,,\qquad\qquad (\rho\to\infty)
\,\,,
\eeq
where $\phi_{\infty}$ is the UV value of the dilaton. For small $\rho$ one gets two possible consistent behaviors, namely  $F\sim \rho^{-{1\over 2}}\,\,,\,\,\rho^{{1\over 2}}$. 
Notice that in one case $F$ diverges at small $\rho$, while in the other
it remains finite.  Actually, when $F$ diverges at $\rho\to 0$ one can show that $F(\rho)$ can be expanded in powers of $\sqrt{\rho}$ as follows:
\beq
F\,=\,{c_0\over \sqrt{\rho}}\,+\,2 N_c\,-\,{N_c^2\over c_0}\,\,\sqrt{\rho}\,+\,
\Big(\,{4\over 3}\,+\,{2N_c^3\over c_0^2}\,\Big)\,\rho+\,\cdots\,\,,
\qquad\qquad (\rho\to\ 0)\,\,,
\label{F-vs-rho-IR}
\eeq
where $c_0$ is a non-zero constant that must be taken to be positive if we want to ensure that $F>0$. Plugging the expansion (\ref{F-vs-rho-IR}) into the right-hand side of (\ref{dilaton-rho}), one can get the IR expansion of the dilaton $\phi$:
\beq
{d\phi\over d\rho}\,=\,-{3N_c\over 4 c_0\sqrt{\rho}}\,+\,{9N_c^2\over 4 c_0^2}\,-\,
{15 N_c^3\over 2 c_0^3}\,\,\sqrt{\rho}\,+\,\cdots\,\,,
\qquad\qquad (\rho\to\ 0)\,\,.
\label{Phi-vs-rho-IR}
\eeq
Notice that $\phi$ is regular as $\rho\to 0$, although $d\phi/d\rho$ diverges. In figure \ref{UnF-Phi} we have plotted the numerical results for $F(\rho)$ and $\phi(\rho)$ for two different values of the constant $c_0$. 

Let us now consider the case in which $F$ is regular as $\rho\to 0$. Apart from the solution in which $F=N_c$ for all $\rho$, there are other solutions where $F$ is not constant and can be expanded near  $\rho\approx 0$ as:
\beq
F\,=\,b_0\,\sqrt{\rho}\,-\,2\Big(\,2\,+\,{b_0^2\over N_c}\,\Big)\,\rho\,+\,
{b_0(5b_0^2\,+\,12 N_c)\over N_c^2}\,\rho^{{3\over 2}}\,+\,\cdots\,\,,
\qquad\qquad (\rho\to\ 0)\,\,.
\eeq
Notice that in this case $F$ vanishes at $\rho\to 0$. By taking $b_0>0$ one can make $F$ positive for small values of $\rho$. However, one can check by numerical integration that, after having a maximum the function $F(\rho)$ starts to decrease and becomes negative as $\rho$ increases. Due to this pathological behavior we will consider this solution as unphysical.

\subsection{Analysis of the general system}
\label{untruncated-unflavored-subsection}
Let us now come back to the general ansatz     (\ref{ansatz})-(\ref{calH}) and
perform an analysis of this  system by using the new radial variable $\rho$ and the function $F$ defined in (\ref{rho}) and (\ref{F}). 
The corresponding BPS equations are written in appendix \ref{BPSapp}.  From eqs. (\ref{hprime})  and (\ref{gprime}) it is easy to verify that the equation that determines the function $F(\rho)$ is given by:
\beq
{dF\over d\rho}\,=\,{A\,\beta+\tilde A\,\tilde\beta\over
D\,\beta+\tilde D\,\tilde\beta}\,\,,
\label{dF}
\eeq
where the functions $A(\rho)$, $\tilde A(\rho)$, $D(\rho)$ and $\tilde D(\rho)$ are:
\bear
&&A\,=\,\Big[\,2\,-\,(1-w^2)\,{F\over 2\rho}\,\Big]\,\Big[\,F\,-\,N_c\,\Big]\,+\,
N_c\,w\,(w-\gamma)\,{F\over \rho}\,\,,\rc\rc
&&\tilde A\,=\,2N_c\,(w-\gamma)\,\sqrt{{F\over \rho}}\,\,,\rc\rc
&&D\,=\,(1-w^2)\,(F+N_c)\,+\,2N_c\,w\,(w-\gamma)\,\,,\rc\rc
&&\tilde D\,=\,{N_c\over 4}\,V\,\sqrt{{F\over \rho}}\,+\,N_c\,
(w-\gamma)\,\sqrt{{\rho\over F}}\,-\,2w\sqrt{F\rho}\,\,\,,
\label{A-D}
\eear
with $V$ being the following function of $w$, $\gamma$ and $\kappa$:
\beq
V\,=\,(w-3\gamma)\,(1-w^2)\,-\,4 w\,+\,8\kappa\,\,.
\label{V-unflavored}
\eeq
The quantities $\beta$ and $\tilde\beta$ appearing in (\ref{dF})  characterize the dependence of the Killing spinors on the holographic coordinate (see appendix \ref{BPSapp}). They can be written in terms of an angle $\alpha$ as $\beta=\cos\alpha$, $\tilde\beta=\sin\alpha$. Alternatively, one can write
$\tan\alpha=\tilde\Lambda/\Lambda$ as in (\ref{betas-Lambdas}). The explicit expressions of $\Lambda$ and $\tilde\Lambda$ are given in (\ref{Lambdas}). In terms of the variables $\rho$ and $F$, they are:
\bear
&&\Lambda\,=\,\rho\,+\,{1-w^2\over 4}\,F\,+\,{N_c\over 4}\,(1+w^2-2 w\gamma)\,-\,
{N_c\,\rho\over 3F}\,\,,\rc\rc
&&\tilde\Lambda\,=\,{N_c\over 24}\,V\,\sqrt{F\over \rho}\,-\,w\sqrt{\rho F}\,+\,
{N_c\over 2}\,(w-\gamma)\,\, \sqrt{{\rho\over F}}\,\,.
\label{Lambdas-rho}
\eear
Similarly, the equations that determine the functions $w(\rho)$ and $\gamma(\rho)$ can be easily obtained from (\ref{BPS-w}) and (\ref{phi-gamma}). Let us write them as:
\beq
{dw\over d\rho}\,=\,{B\beta+\tilde B\tilde\beta\over
D\beta+\tilde D\tilde\beta}\,\,,
\qquad\qquad
{d\gamma\over d\rho}\,=\,{C\beta+\tilde C\tilde\beta\over
D\beta+\tilde D\tilde\beta}\,\,,
\label{w-gamma-rho}
\eeq
where $D(\rho)$ and  $\tilde D(\rho)$ are the same as in (\ref{A-D}) and the new functions 
$B(\rho)$, $\tilde B(\rho)$, $C(\rho)$ and $\tilde C(\rho)$ are:
\bear
&&B\,=\,{2N_c\over 3}\,\Bigg[\,{V\over 4\rho}\,-\,3\,{w-\gamma\over F}\,\Bigg]\,\,,\rc\rc
&&\tilde B\,=\,{4\over 3}\,\Bigg[\,\Big(\,3\,-\,{2N_c\over F}\,\Big)\,\sqrt{{\rho\over F}}\,-\,
{3\over 4}(1-w^2)\,\sqrt{{F\over \rho}}\,\,\Bigg]\,\,,\rc\rc
&&C\,=\,{2\over 3}\,\Bigg[\,{V\over 4\rho}\,F\,+\,3(w-\gamma)\,\Bigg]\,\,,\rc\rc
&&\tilde C\,=\,{4\over 3}\,\Bigg[\,\sqrt{{\rho\over F}}\,-\,{3\over 4}\,\,
(1+w^2-2w\gamma)\,\sqrt{{F\over \rho}}\,\,\Bigg]\,\,.
\label{BC-unflavor}
\eear
Moreover, from (\ref{hprime}) one easily gets that the jacobian for the change of the radial variable is:
\beq
{dr\over d\rho}\,=\,{\sqrt{F}\over D\beta+\tilde D\tilde\beta}\,\,.
\label{drdho-untruncated}
\eeq
In terms of these quantities, the metric can be written as:
\beq
ds^2\,=\,e^{{\phi\over 2}}\,\Bigg[\,dx^2_{1,2}\,+\,
{F\over \big(\,D\beta+\tilde D\tilde\beta\,\big)^2}\,(d\rho)^2\,+\,
{\rho\over 4}\,(\sigma^i)^2\,+\,{F\over 4}\,\Big(\,\omega^i\,-\,A^i\,\Big)^2
\,\Bigg]\,\,.
\label{metric-in-rho}
\eeq
Similarly, from (\ref{phi-gamma}) one can obtain the differential equation that determines the dependence of the dilaton $\phi$ on the variable $\rho$, namely:
\beq
{d\phi\over d\rho}\,=\,{E\beta+\tilde E\tilde\beta\over
D\beta+\tilde D\tilde\beta}\,\,,
\label{dilaton-in-rho}
\eeq
where the new functions $E(\rho)$ and $\tilde E(\rho)$ are given by:
\bear
&&E\,=\,N_c\,\Big[\,{1\over F}\,-\,{3(1+w^2-2w\gamma)\over 4\rho}\,\Big]\,\,,\rc\rc
&&\tilde E\,=\,-N_c\,\Big[\,{V\over 8\rho}\,\,\sqrt{F\over \rho}\,+\,
{3(w-\gamma)\over 2\sqrt{F\rho}}\,\Big]\,\,.
\label{E-tildeE}
\eear
As a check of eqs. (\ref{dF})-(\ref{E-tildeE}) one can verify that they reduce to the ones of the truncated system when $w=\gamma=\kappa=0$. Notice that in this case $\beta=1$ and $\tilde\beta=V=0$ and, as a consequence $B$ and $C$ vanish and eq. (\ref{w-gamma-rho}) is solved by the truncated values (\ref{truncation}). Moreover, one easily demonstrates that, in this case, (\ref{dF}) and (\ref{dilaton-in-rho}) reduces to (\ref{F-rho})  and  (\ref{dilaton-rho}) respectively. 

\subsubsection{Initial conditions}

Given a set of initial conditions for the functions $F$, $w$, $\gamma$ and $\phi$, and a value of the integration constant $\kappa$, the system of equations (\ref{dF}), (\ref{w-gamma-rho}) and (\ref{dilaton-in-rho}) can be numerically integrated. Let us see how one can determine these initial data in a meaningful  way.  First of all, let us fix the value of the function $w(\rho)$ at $\rho=0$. Recall (see (\ref{Ai-w})) that $w$ parameterizes the one-form $A^i$ which, in turn, determines the mixing of the two three-spheres in the ten-dimensional fibered geometry. The curvature of the gauge connection $A^i$ (defined as in (\ref{Fi}) with $B^i\to A^i$) determines the non-triviality of this mixing. Indeed, if it vanishes the one-forms $A^i$ are a pure gauge connection that can be taken to vanish after a suitable gauge transformation. In this case one can choose a new set of three one-forms in which 
the two three-spheres are disentangled in a manifest way. On the other hand, from the wrapped brane origin of our solutions, one naturally expects such an un-mixing of the two $S^3$'s  to occur in the IR limit $\rho=0$ of the metric, where it should be possible to factorize the directions parallel and orthogonal to the brane worldvolume in a well-defined way. Moreover, by  a direct calculation using (\ref{sigma-w}) it is easy to verify that for $w=1$ the curvature of the  one-form $A^i$ vanishes and, thus, $A^i$ is pure gauge. Thus, it follows that the natural initial condition for $w(\rho)$ is:
\beq
w(\rho=0)\,=\,1\,\,.
\label{initialw}
\eeq

Let us now fix the value of the constant $\kappa$ by adapting the procedure employed in ref. \cite{Maldacena:2001pb} in the case of backgrounds that are obtained by uplifting from seven-dimensional gauged supergravity.  In this reference the authors determined $\kappa$ by imposing the vanishing at the origin of the pullback  of the RR three-form on the three-cycle  of the seven dimensional geometry which, in our notations, is the one parameterized by the one-forms $\sigma^i$. In the seven dimensional approach
this three cycle  shrinks at the origin and can be naturally interpreted as the one on which  the fivebranes are wrapped. This procedure is possibly ambiguous when one tries to apply it in the ten-dimensional geometry, where actual D5-branes live. Moreover, the solutions  studied here cannot be obtained, in general, by uplifting from seven dimensions.  Therefore, it is convenient to search for a way to fix $\kappa$ directly in ten dimensions.

 We start  by noting that the seven dimensional cycle, parametrized by the one-forms $\sigma^i$, does not shrink in the ten dimensional geometry and, thus, it does not look strictly necessary that the RR three-form flux vanishes on it. Indeed, it does not shrink even in the solutions found in \cite{Maldacena:2001pb}. We  think that the relevant cycle, which should also be the cycle on which the branes are wrapped, is:
 \beq
 \Sigma\equiv\{\omega^i=\sigma^i\}.
 \label{mixc}
 \eeq
To understand this, let us  begin by pointing out  that, even if the seven dimensional gauge field $A^i$ is pure gauge at the origin when the initial condition (\ref{initialw}) holds, it is not vanishing there. This non-vanishing of the gauge connection is the origin of the mixing among the two three cycles in the ten-dimensional fibered geometry. As we are going to argue, this mixing is taken into account if one considers the cycle (\ref{mixc})
\footnote{Alternatively, by performing a gauge transformation to $A^i$ one can get a new gauge connection $\tilde A^i={1-w\over 2}\,\tilde \sigma^i$, where $\tilde \sigma^i$ is a new set of left-invariant one-forms. In this new gauge the condition (\ref{initialw}) implies that $\tilde A^i$ vanishes at the origin and the analogue of the cycle $\Sigma$ is just the cycle parameterized by the $\tilde \sigma^i$'s with $w^i=0$.
}. It is indeed easy to see that that cycle $\Sigma$ is actually shrinking in the full ten-dimensional geometry if some regularity conditions are satisfied. Let us require that the  metric function $F(\rho)$ approaches a constant finite value $F_0$ as $\rho\to 0$, namely:
\beq
F\approx F_0\,\,,
\qquad\qquad (\rho\sim 0)\,\,.
\label{F-IR}
\eeq
The induced metric on $\Sigma$ is:
\beq
ds^2_{\Sigma}\,=\,{e^{{\phi\over 2}}\over 4}\,\,\Big[\,\rho\,+\,{(1-w)^2\over 4}\,\,F\,\,
\Big]\,(\sigma^i)^2\,\,.
\label{ds-on-Sigma}
\eeq
Obviously, due to the factor in brackets in (\ref{ds-on-Sigma}), $ds^2_{\Sigma}\to 0$ as 
$\rho\to 0$ if eqs. (\ref{initialw}) and (\ref{F-IR}) hold and the dilaton $\phi$ is finite at the origin. Moreover, in order to have a non-singular RR flux at the origin, one should require that $F_3$ vanishes on $\Sigma$ when $\rho\to 0$. We take this condition as a general criterium to fix the value of $\kappa$.  Remarkably, as can be easily verified from our ansatz, the pullback of $F_3$ on $\Sigma$ is independent of $\rho$ and given by:
\beq
{F_3}{\big|_\Sigma}\,=\,{N_c\over 4}\,\Big(\,\kappa-{1\over 2}\,\Big)\,
\sigma^1\wedge\sigma^2\wedge \sigma^3\,\,.
\eeq
Therefore, it is clear that we must fix the value of the constant $\kappa$ to the value:
\beq
\kappa={1\over 2}\,\,.
\label{kappa-onehalf}
\eeq
Notice that this is exactly the value of $\kappa$ used in \cite{Maldacena:2001pb}. 
Let us see how one can reobtain this same value of $\kappa$ by requiring that the dilaton is finite at $\rho=0$.  Let $V_0$ be the value of the function $V$ defined  in 
(\ref{V-unflavored}) at $\rho=0$. Let us assume that $V_0\not=0$ and that $w$ and $F$ satisfy the initial conditions (\ref{initialw}) and (\ref{F-IR}). Then, by inspecting (\ref{Lambdas-rho}) one concludes that $\tilde\Lambda$ diverges at $\rho\sim 0$:
\beq
\tilde \Lambda\,\approx\, {N_c\over 24}\,\,V_0\,\sqrt{F_0}\,\,{1\over \sqrt{ \rho}}\,\,,
\qquad\qquad (\rho\sim 0)\,\,,
\label{tildeLambda-IR}
\eeq
while $\Lambda$ remains finite at $\rho= 0$.  This means that $\beta\approx 0$ as $\rho\to 0$ and, therefore, the differential equation (\ref{dilaton-in-rho}) for the dilaton reduces approximately to:
\beq
{d\phi\over d\rho}\approx{\tilde E\over \tilde D}\,\,,
\qquad\qquad (\rho\sim 0)\,\,.
\label{IRdilaton}
\eeq
Moreover, from (\ref{E-tildeE}) and  (\ref{A-D}) we get that the leading behavior of the coefficients $\tilde E$ and $\tilde D$ as $\rho\to 0$ is:
\beq
\tilde E\approx -N_c\,{ V_0\over 8\rho^{{3\over 2}}}\,\sqrt{F_0}\,\,,
\qquad\qquad
\tilde D\approx {N_c\over 4}\, V_0\,{\sqrt{F_0}\over \rho^{{1\over 2}}}\,\,,
\qquad\qquad (\rho\approx 0)\,\,.
\eeq
Therefore, the first-order equation (\ref{IRdilaton})  for the dilaton becomes:
\beq
{d\phi\over d\rho}\approx \,-{1\over 2\rho}\,\,,
\qquad\qquad (\rho\approx 0)\,\,,
\eeq
which, upon integration, gives rise to the divergent IR behaviour: 
\beq
\phi\,\sim\,-{1\over 2}\,\log\rho\,+\,o(\rho)\,\,.
\label{divergent-dilaton}
\eeq
The only way to escape this conclusion is by requiring the vanishing of $V_0$, namely:
\beq
V_0\,=\,0\,\,.
\label{V0=0}
\eeq
But, from the expression for $V$ in (\ref{V-unflavored}), we get that:
\beq
V_0\,=\,8\Big(\kappa-{1\over 2}\Big)\,\,,
\eeq
and, thus, the condition (\ref{V0=0})  fixes again  the value of the constant $\kappa$ to that written in eq. (\ref{kappa-onehalf}). Notice that, contrary to what happens in (\ref{tildeLambda-IR}),  $\tilde\Lambda$ does not diverge at $\rho=0$ when $V_0\,=\,0$. Actually, $\tilde\Lambda\to 0$ in this case and, therefore, the only possibility of having $\beta\approx 0$ for $\rho\to 0$,  as is required to deduce  (\ref{IRdilaton}),  is by imposing that $\Lambda$ vanishes faster than $\tilde \Lambda$ as $\rho\to 0$ which, in particular, implies that we must require:
\beq
\Lambda(\rho=0)\,=\,0\,\,.
\label{Lambda=0}
\eeq
 If, on the contrary, (\ref{Lambda=0}) is not satisfied, one has that $\tilde\beta\approx 0$ as $\rho\to 0$ and  $d\phi/d\rho\approx E/D\sim 1/\rho$, which, again,  gives rise to the undesired behavior  $\phi\sim \log \rho$ near $\rho\approx 0$. Thus, in order to have a regular dilaton at $\rho=0$, we should impose  the condition (\ref{Lambda=0}). Actually, 
 from the expression of $\Lambda$ in (\ref{Lambdas-rho}), as well as the initial conditions (\ref{initialw}) and (\ref{F-IR}), it is immediately possible to conclude that (\ref{Lambda=0}) implies that the IR value of $\gamma$ should be fine-tuned to the value:
\beq
\gamma(\rho=0)\,=\,1\,\,.
\label{gamma=0}
\eeq
If (\ref{gamma=0}) holds, equation (\ref{IRdilaton}) is still valid and one can check that, indeed, the dilaton remains finite in the IR.

It is also interesting to look at the IR form of the metric (\ref{metric-in-rho}) when the initial conditions just found are satisfied. Since in this case $\beta\to 0$, only the behavior of $\tilde D$ near $\rho\to 0$ is relevant. One has:
 \beq
\tilde D\,\approx\,-2\sqrt{F_0}\,\rho^{{1\over 2}}\,\,,
\qquad\qquad (\rho\approx 0)\,\,.
\eeq
Using this result in (\ref{metric-in-rho}), one gets that 
the $(\rho, \sigma^i)$ part of the metric near $\rho\sim 0$ takes the form:
\beq
{\,d\rho^2\over 4\rho}\,+\,{\rho\over 4}\,\,
(\sigma^i)^2\,\,.
\label{rho-sigma-metric}
\eeq
Let us now change the radial variable to:
\beq
\rho=\tau^2\,\,.
\eeq
The resulting metric in the $(\tau, \sigma^i)$ sector is:
\beq
d\tau^2+{\tau^2\over 4}\,\,
(\sigma^i)^2\,\,,
\eeq
which is just the metric of  flat four-dimensional Euclidean space. Thus, one expects that the metric for these solutions is regular at $\tau=0$. We have verified this fact by explicitly computing the scalar curvature  for our solutions and by checking that it remains finite at $\tau=0$.

\subsubsection{Explicit solution}
\label{explicitsolution-unflavored-section}

Let us now solve the BPS equations in a series expansion around $\rho\approx 0$. For this purpose, let us suppose that $F(\rho)$ is given by the series:
\beq
F\,=\, F_0\,+\,F_1\,\rho\,+\,F_2\,\rho^2\,+\,\cdots\,\,.
\eeq
Then, by plugging this expansion into the BPS equations one can get the coefficients
$F_n$ for $n\ge 1$ in terms of $F_0$. The corresponding expression of $F_1$ and $F_2$ is:
\bear
F_1&=&{(F_0-N_c)(9F_0+5N_c)\over 12 F_0^2}\,\,,\rc\rc
F_2&=&{(F_0-N_c)(\,36F_0^3\,-\,4F_0^2 N_c\,+\,19F_0 N_c^2\,+\,23 N_c^3\,)\over
144 F_0^5}\,\,.
\eear

Interestingly, one can verify that  when $F_0=N_c$ the coefficients $F_n$ vanish for $n\ge 1$ and the exact solution is $F=N_c$ as in the background studied in \cite{Maldacena:2001pb}. Similarly, for the initial conditions at $\rho=0$  displayed in (\ref{initialw}) and (\ref{gamma=0}), the functions $w(\rho)$ and $\gamma(\rho)$ can be written as:
\bear
w&=&1\,+\,w_1\,\rho\,+\,w_2\,\rho^2\,+\,\cdots\,\,,\rc\rc
\gamma&=&1\,+\,\gamma_1\,\rho\,+\,\gamma_2\,\rho^2\,+\,\cdots\,\,,
\eear
with the first two coefficients given by:
\bear
&&w_1\,=\,{2N_c-3F_0\over 3F_0^2}\,\,,
\qquad\qquad
w_2\,=\,{18N_c^3\,-\,19\,F_0 N_c^2\,-\,16 F_0^2\,N_c\,+\,18 F_0^3\over 
36 F_0^5}\,\,,\rc\rc
&&\gamma_1\,=\,-{1\over 3F_0}\,\,,
\qquad\qquad\qquad\,\,
\gamma_2\,=\,-{4N_c^2-4F_0 N_c - F_0^2\over 36 F_0^2}\,\,.
\eear

\begin{figure}[ht]
\begin{center}
\includegraphics[width=0.95\textwidth]{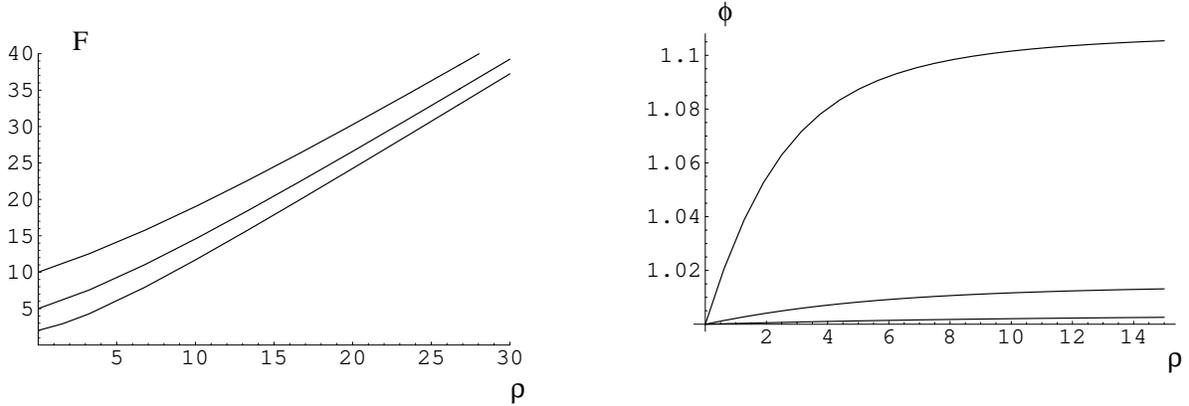}
\end{center}
\caption[Unflavor-F-Phi]{$F$ and $\phi$ for three different values of the initial condition $F_0$ ($F_0=2, 5 $ and $10$) for $N_c=1$. } 
\label{F-Phi-unflavor}
\end{figure}

One can verify that for $F_0=N_c$ one has $w_n=\gamma_n$ for all  values of the index $n$. Indeed, in this case our generalized solution collapses to the solution studied in \cite{Maldacena:2001pb}. Moreover, the functions $\beta$ and $\tilde\beta$ behave near $\rho\approx 0$ as:
\beq
\beta\,=\,{1\over 2 \sqrt{F_0}}\,\Big(3-{N_c\over F_0}\Big)\,\sqrt{\rho}\,+\,
o(\rho^{{3\over 2}})\,\,,
\qquad\qquad
\tilde\beta\,=\,-1\,+\,o(\rho)\,\,,
\eeq
while, for $\rho\to 0$ the dilaton can be expanded as:
\beq
\phi\,=\,\phi_0\,+\,{7N_c^2\over 24 F_0^3}\,\rho\,+\,o(\rho^2)\,\,,
\label{IR-dilaton-untruncated-noflavor}
\eeq
where $\phi_0$ is an integration constant. In particular this result implies that $\phi$ is regular at $\rho=0$, as claimed above. 

\begin{figure}[ht]
\begin{center}
\includegraphics[width=0.95\textwidth]{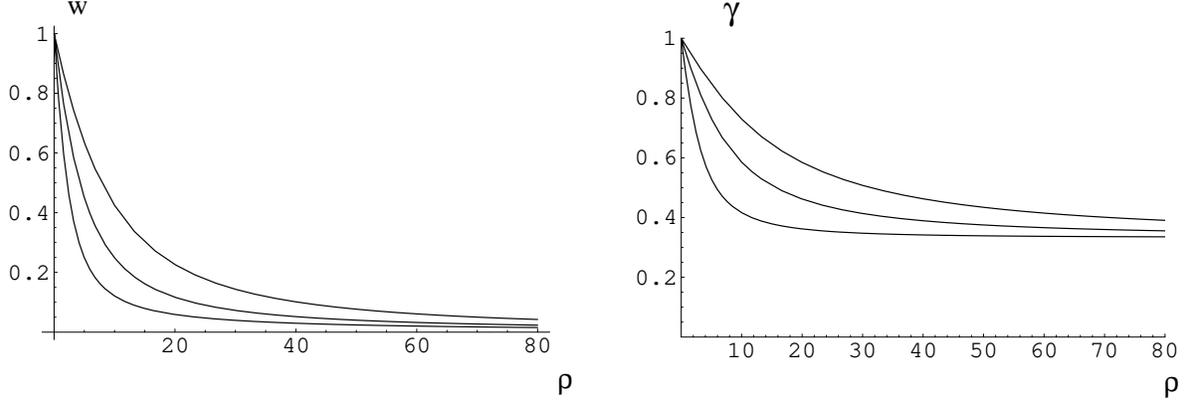}
\end{center}
\caption[Unflavor-w-gamma]{The functions $w$ and $\gamma$ for three different values of the initial condition $F_0$ ($F_0=2, 5 $ and $10$) and $N_c=1$.} 
\label{w-gamma-unflavor}
\end{figure}

The system of BPS equations can be solved numerically with the initial conditions  just found.  From this numerical analysis we notice that, in addition to the solutions analyzed in \cite{Maldacena:2001pb} (for which $F=N_c$ and $w=\gamma$) there are others  which, for $\rho\to\infty$, behave as:
\beq
F\to\infty\,\,,\qquad\qquad
w\to 0\,\,,\qquad\qquad
\gamma\to\gamma_*\,\,,
\qquad\qquad
(\rho\to\infty)\,\,,
\label{UV-unflavor}
\eeq
where $\gamma_*$ is  a finite value. In figure \ref{F-Phi-unflavor} we have plotted the function $F$ and the dilaton for several values of the constant $F_0$. These curves should be compared with the ones in figure \ref{UnF-Phi}. The main differences are in the IR behavior of $F(\rho)$, which is now finite at $\rho=0$.  In all these solutions the dilaton $\phi$ is asymptotically constant in the UV, in contrast with the ones 
of \cite{Maldacena:2001pb}, for which the dilaton grows linearly with the holographic coordinate. In figure \ref{w-gamma-unflavor} we have represented the functions $w$ and $\gamma$ for the same set of values of $F_0$ as in figure \ref{F-Phi-unflavor}. 

The behaviour (\ref{UV-unflavor}) is easy to reproduce analytically by studying the system of the BPS equations. Indeed, if $F$, $w$ and $\gamma$ behave as in 
(\ref{UV-unflavor}) then one readily gets  from (\ref{Lambdas-rho}) that, at leading order,  $\Lambda\to\infty$ and $\tilde\Lambda\to {\rm constant}$  for large $\rho$ and, as a consequence,   $\beta\to 1$ and $\tilde\beta\to 0$  as $\rho\to\infty$. Moreover, one can straightforwardly demonstrate that 
equation (\ref{dF}) determining $F$ reduces to the one found in the truncated system in
(\ref{FNc=0}). From the general solution written in (\ref{FNc=0-sol}) we see that 
$F\approx 4\rho /3$ for $\rho\to\infty$. Furthermore, one can verify that
 it is consistent to take the following behavior of $w$ as $\rho\to\infty$:
\beq
w\approx {b\over \rho}\,\,,
\label{asymp-w}
\eeq
with $b$ being a constant to be determined. Then, at leading order,  one gets from (\ref{Lambdas-rho}):
\bear
&&\Lambda\,\approx \,{4\over 3}\,\rho\,\,,\rc\rc
&&\tilde\Lambda\,\approx\,{N_c\over 12\sqrt{3}}\,V_*\,-\,{2b\over \sqrt{3}}\,
-\,{\sqrt{3}\over 4}\,N_c\,\gamma_*
\,\equiv \,\tilde\Lambda_*\,\,,
\qquad\qquad
(\rho\to\infty)\,\,,
\eear
where $V_*$ is the asymptotic value of $V$ at $\rho\to\infty$. Taking into account the expression of $V$ (eq. (\ref{V-unflavored})), this value can be written in terms of the asymptotic value of $\gamma$ as follows:
\beq
V_*\,=\,8\kappa\,-\,3\gamma_*\,\,,
\label{V_*}
\eeq
where we have momentarily considered a general value of the integration
constant $\kappa$.
Notice also  that $\beta$ and $\tilde\beta$ behave as:
\bear
&&\beta\,\approx\,1\,-\,{\tilde\Lambda_*^2\over 2\Lambda^2}\,
\approx\,1\,-\,{9\tilde\Lambda_*^2\over 32}\,\,{1\over \rho^2}\,\,,\rc\rc
&&\tilde\beta\,\approx\,{\tilde\Lambda\over\Lambda}\,
\approx\,{3\tilde\Lambda_*\over 4}\,{1\over \rho}\,\,,
\qquad\qquad\qquad\qquad (\rho\to\infty)\,\,.
\label{asymp-betas}
\eear
Using this result one can write asymptotically the differential equation for $\gamma(\rho)$ as:
\beq
{d\gamma\over d\rho}\,\approx\,{C\over D}\,\approx\,
\Big[\,{V_*\over 3}\,-\,3\gamma_*\,\Big]\,{1\over2 \rho}\,\,.
\eeq
Consistency at leading order requires that $\gamma_*$ and $V_*$ must be related by:
\beq
\gamma_*\,=\,{V_*\over 9}\,\,.
\eeq
Taking into account the value of $V_*$ written in (\ref{V_*}), one gets the value of $\gamma_*$ in terms of the constant $\kappa$, namely:
\beq
\gamma_*\,=\,{2\kappa\over 3}\,\,.
\label{gamma*}
\eeq
Notice then that the asymptotic value of $V$ is:
\beq
V_*\,=\,6\kappa\,\,.
\eeq
Let us now calculate the coefficient $b$ that determines the asymptotic behavior of the function $w$. With this purpose, we notice that the functions $B$ and $\tilde B$ defined in (\ref{BC-unflavor})  behave as:
\bear
&&B\,=\,{N_c\over 6\rho}\,\Big[\,V_*\,+\,9\gamma_*\,\Big]\,\,,\rc\rc
&&\tilde B\,=\,{4\over \sqrt{3}}\,\,,
\qquad\qquad\qquad\qquad (\rho\to\infty)\,\,.
\eear
Also taking into account that $F\sim 4\rho/3$,   as  well as eqs. (\ref{asymp-w}) and (\ref{asymp-betas}), one gets that:
\beq
{dw\over d\rho}\,=\,{3\over 2\rho^2}\,\Big[\,{N_c\over 8}\,\,
(\,V_*\,+\,3\gamma_*\,)\,-\,b\,\Big]\,\,,
\eeq
which, for consistency with  (\ref{asymp-w}), implies that:
\beq
b\,=\,3N_c\,\kappa\,\,.
\eeq
Finally, 
one can verify from (\ref{dilaton-in-rho})  that the dilaton  $\phi$ reaches a constant value when  $\rho\to\infty$.

Taking into account that the regularity conditions in the IR fix $\kappa$ to be equal to $1/2$, one gets that the actual values of $\gamma_*$ and $b$ are:
\beq
\gamma_*\,=\,{1\over 3}\,\,,
\qquad\qquad
b\,=\,{3N_c\over 2}\,\,,
\eeq
a result which is confirmed by our numerical solutions. 

It is interesting to compare the background found here with the one obtained in 
\cite{Maldacena:2001pb}. The latter corresponds to D5-branes wrapped on a three-cycle of a manifold of $G_2$ holonomy in  the near-horizon limit which, as it should, has a dilaton which grows linearly with the holographic coordinate in the UV. In our case the dilaton is asymptotically constant and the metric approaches that of a $G_2$ cone as we move towards the UV, while in the IR region our solution is qualitatively similar to the one analyzed in \cite{Maldacena:2001pb}. It is thus natural to regard our solution as corresponding to D5-branes wrapped on a three-cycle of a $G_2$ cone, in which the near horizon limit has not been taken and, thus, as we move towards the large $\rho$ region the effect of the branes on the metric becomes asymptotically negligible and we recover the  geometry of the $G_2$ cone where the branes are wrapped. Notice that in reference \cite{Casero:2006pt} the authors found similar backgrounds for the case of D5-branes wrapped on a two-cycle. In this case the solutions asymptotically approach the conifold geometry.

\setcounter{equation}{0}

\section{Addition of flavor}
\label{flavor-addition-section}

Our main motivation to study a generalized ansatz of the form (\ref{ansatz})-(\ref{F3ansatz}) was to explore the addition of unquenched flavors to the supergravity duals of ${\cal N}=1$ supersymmetric gauge theories in 2+1 dimensions. Indeed, we will show below that the backreacted flavored metrics that we will find can be represented in the form (\ref{ansatz}), \ie\ their deformation with respect to the unflavored ones of \cite{Maldacena:2001pb}  is of just the type studied in section \ref{Unflavor-section}.  
We will achieve this conclusion in three steps. First of all, we will study the problem in the approximation in which the flavor brane is considered as a probe in the unflavored background. 

The appropriate flavor branes for our case are wrapped D5-branes that fill the Minkowski   spacetime and are extended in the holographic direction. By using kappa symmetry \cite{swedes} of the probe we will be able to find some simple configurations that preserve all supersymmetries of the background. In these configurations the D5-branes are extended along a submanifold of the internal space that has the topology of a cylinder and reaches the origin of the holographic coordinate. They can be used to add massless flavors to the gravity dual of \cite{Maldacena:2001pb}.  Actually there is a continuous family of such supersymmetry preserving  embeddings.  In a second step we will determine how to combine these embeddings to produce a distribution of them that produces a backreaction on the background such that the metric is still of the form (\ref{ansatz}). In general the flavor branes act as sources for the RR fields and also modify the energy-momentum tensor. Due to the fact that we will consider a continuous distribution of D5-branes, these extra terms are not localized and, as we will see, their influence in the background can be obtained. 

The main modification of the backreacted ansatz with respect to the one studied in section \ref{Unflavor-section} is that in the unquenched case the new RR source terms give rise to a violation of the Bianchi identity for $F_3$. In a third step we will determine the appropriate modification of $F_3$ that gives rise to the desired violation of the Bianchi identity. Moreover, once $F_3$ is known we can use it in the supersymmetry variations and obtain the BPS equations of the flavored backgrounds, exactly in the same way as in the unflavored system. This analysis is performed in appendix \ref{BPSapp}, whereas the study of the different solutions of the BPS system will be carried out in the remaining sections of this paper. In appendix \ref{EOMapp} we show that the backgrounds obtained in this way solve the second order equations of motion of the supergravity plus brane system.

\subsection{Supersymmetric probes}

Let us consider a D5-brane probe in some of the  backgrounds studied in section \ref{Unflavor-section} and let $\xi^{\mu}$ ($\mu=0,\cdots ,5$) be a set of worldvolume  coordinates. If $X^{M}$ denote ten-dimensional coordinates,  the D5-brane embedding will be characterized by a set
of functions  $X^{M}(\xi^{\mu})$. The induced metric on the worldvolume is:
\beq
\hat G_{\mu\nu}^{(6)}\,=\,\partial_{\mu} X^{M}\,\partial_{\nu} X^{N}\,G_{MN}\,\,,
\label{inducedmetric}
\eeq
where $G_{MN}$ is the ten-dimensional metric.  The embeddings of the D5-brane probe that preserve the supersymmetry of the background are those that satisfy the kappa symmetry condition \cite{bbs}:
\beq
\Gamma_{\kappa}\,\epsilon\,=\,\epsilon\,\,,
\eeq
where $\Gamma_{\kappa}$ is a matrix that depends on the embedding of the probe and $\epsilon$ is a Killing spinor of the background. Acting on spinors $\epsilon$ such that $\epsilon=i\epsilon^*$ (as the ones of our background, see (\ref{projections})) and assuming that there is no worldvolume gauge field, the matrix $\Gamma_{\kappa}$ for a D5-brane probe  is \cite{swedes,bbs}:
\beq
\Gamma_{\kappa}\,=\,{1\over 6!}\,\,{1\over \sqrt { -\hat G_6}}\,\,\,
\epsilon^{\mu_1\cdots \mu_6}\,\,\gamma_{\mu_1\cdots \mu_6}\,\,,
\label{kappamatrix}
\eeq
where $\hat G_6$ is the determinant of the induced metric $\hat G_{\mu\nu}^{(6)}$ and
$\gamma_{\mu_1\cdots \mu_6}$ is the antisymmetrized product of worldvolume Dirac  matrices $\gamma_{\mu}$. In order to define these induced matrices, let us denote by
$E_{N}^{\underline{M}}$ the coefficients that appear in the expression of  the
frame one-forms $e^{\underline{M}}$ of the ten-dimensional metric  in terms of
the differentials of the coordinates, namely:
\beq
e^{\underline{M}}\,=\,E_{N}^{\underline{M}}\,dX^N\,\,.
\eeq
Then, the induced Dirac matrices on the worldvolume are defined as
\beq
\gamma_{\mu}\,=\,\partial_{\mu}\,X^{M}\,E_{M}^{\underline{N}}\,\,
\Gamma_{\underline{N}}\,\,,
\label{wvgamma}
\eeq
where $\Gamma_{\underline{N}}$ are constant ten-dimensional Dirac matrices.
Moreover, the pullback of the frame one-forms $e^{\underline{M}}$ is given by:
\beq
P[\,e^{\underline{M}}\,]\,=\,E_{N}^{\underline{M}}
\partial_{\mu}\,X^{N}\,d\xi^{\mu}\,\equiv C_{\mu}^{\underline{M}}
\,d\xi^{\mu}\,\,,
\eeq
where, in the last step,  we have defined the coefficients 
$C_{\mu}^{\underline{M}}\equiv E_{N}^{\underline{M}}\partial_{\mu}\,X^{N}$.
Notice that  the induced Dirac matrices $\gamma_{\mu}$ can be expressed in
terms of the constant $\Gamma$'s by means of these same coefficients 
$C_{\mu}^{\underline{M}}$, namely:
\beq
\gamma_{\mu}\,=\,
C_{\mu}^{\underline{M}}\,\Gamma_{\underline{M}}\,\,.
\label{inducedgamma}
\eeq
In order to obtain the particular class of D5-brane embeddings that we are interested in to add flavor to the supergravity dual of 2+1 dimensional gauge theories, let  us  parameterize  the forms $\sigma^i$ and $\omega^i$  in terms of the angles 
$\theta_i$, $\varphi_i$ and $\psi_i$ $(i=1,2)$ as follows:
\bear
\sigma^1&=& \cos\psi_1 d\theta_1\,+\,\sin\psi_1\sin\theta_1
d\varphi_1\,\,,
\qquad\qquad\,\,\,
\omega^1= \cos\psi_2 d\theta_2\,+\,\sin\psi_2\sin\theta_2
d\varphi_2\,\,,\rc
\sigma^2&=&-\sin\psi_1 d\theta_1\,+\,\cos\psi_1\sin\theta_1
d\varphi_1\,\,,\qquad\quad\,\,\,
\omega^2=-\sin\psi_2 d\theta_2\,+\,\cos\psi_2\sin\theta_2 
d\varphi_2\,\,,\
\rc
\sigma^3&=&d\psi_1\,+\,\cos\theta_1 d\varphi_1\,\,,
\qquad\qquad\qquad\qquad\,\,\,\,\,\,\,\,
\omega^3=d\psi_2\,+\,\cos\theta_2 d\varphi_2\,\,.
\eear
Next,  let us consider a D5-brane probe with worldvolume coordinates:
\beq
\xi^{\mu}\,=\,(x^0,x^1,x^2, \psi_1, r, \psi_2)\,\,,
\label{wv-cordinates}
\eeq
and let us embed the D5-brane in the general geometry in such a way that:
\beq
\theta_i\,=\,{\rm constant}\,\,,
\qquad
\varphi_i\,=\,{\rm constant}\,\,,
\qquad
(i=1,2)\,\,.
\label{embeddings}
\eeq
Let us choose the same vierbein basis as in (\ref{basis}). Then, 
for the embedding (\ref{embeddings})  the pullbacks of the frame one-forms are:
\bear
&&P\big[\,e^{x^i}\,\big]\,=\,e^{f}\,dx^i\,\,,
\,\,\,\,\,\,\,\,\,\,\,\,\,\,
(i=0,1,2)\,\,,
\,\,\,\,\,\,\,\,\,\,\,\,\,\,\,\,\,\,\,\,\,\,\,\,\,\,\,\,
P\big[\,e^r\,\big]\,=\,e^{f}\,dr\,\,,\rc\rc
&&P\big[\,e^{1}\,\big]\,=\,P\big[\,e^{2}\,\big]\,=\,0\,\,,
\qquad\qquad
P\big[\,e^{3}\,\big]\,=\,{e^{f+h}\over 2}
\,\,d\psi_1\,\,,\rc\rc &&P\big[\,e^{\hat 1}\,\big]\,=\,P\big[\,e^{\hat
2}\,\big]\,=\,0\,\,,
\qquad\qquad
P\big[\,e^{\hat 3}\,\big]\,=\,{e^{f+g}\over 2}\,
\big[\,d\psi_2\,-\,{1+w \over 2}\,d\psi_1\,\big]\,\,.
\eear
Therefore, the induced $\gamma$-matrices are:
\bear
&&\gamma_{x^i}\,=\,e^{f}\,\Gamma_{x^i}\,\,,
\,\,\,\,\,\,\,\,\,\,\,\,\,\,
(i=0,1,2)\,\,,
\,\,\,\,\,\,\,\,\,\,\,\,\,\,\,\,\,\,\,\,\,\,\,\,\,\,\,\,
\gamma_{r}\,=\,e^{f}\,\Gamma_r\,\,,\rc\rc
&&\gamma_{\psi_1}\,=\,{e^{f}\over 2}\,
\big(\,e^{h}\,\,\Gamma_3\,-\,{1+w \over 2}\,\,
e^{g}\,\Gamma_{\hat 3}\,\big)\,\,,
\qquad\qquad\quad
\gamma_{\psi_2}\,=\,{e^{f+g}\over 2}\,\Gamma_{\hat 3}\,\,,
\eear
and, thus, the kappa symmetry matrix $\Gamma_\kappa$  of eq. (\ref{kappamatrix})
is:
\beq
\Gamma_\kappa\,=\,{1\over \sqrt{-\hat G_6}}\,\,
\gamma_{x^0x^1x^2\psi_1 r\psi_2}\,\,.
\eeq
By using the expression of the induced Dirac matrices written above, we get:
\beq
\Gamma_\kappa\,\epsilon\,=\,
{e^{6f+h+g}\over 4\sqrt{- \hat G_6}}\,\,
\Gamma_{x^0x^1x^2}\,\Gamma_3\,\Gamma_r\Gamma_{\hat 3}\,\epsilon\,\,.
\eeq
Moreover, in the type IIB theory the total ten-dimensional chirality of the spinors is fixed. Thus:
\beq
\Gamma_{x^0x^1x^2}\,\Gamma_{123}\,\Gamma_r\,\hat \Gamma_{123}\,\epsilon\,=\,
\epsilon\,\,.
\eeq
Taking into account that $\Gamma_{12}\hat \Gamma_{12}\epsilon\,=\,\epsilon$
(see eq. (\ref{projections})), we conclude that:
\beq
\Gamma_{x^0x^1x^2}\,\Gamma_3\,\Gamma_r\Gamma_{\hat 3}\,\epsilon\,=\,\epsilon\,\,.
\eeq
Moreover, by computing the determinant of the induced geometry for these
embeddings, we arrive at:
\beq
\sqrt{-\hat G_6}\,=\,{e^{6f+h+g}\over 4}\,\,.
\eeq
From the last two equations it follows that, indeed,  $\Gamma_{\kappa}\,\epsilon\,=\,\epsilon$,  \ie\ these cylinder embeddings preserve all supersymmetries of the background. 

\subsection{Smeared configurations}

Notice that the embeddings just considered are mutually supersymmetric for any value of the transverse angles $\theta_1$, $\varphi_1$, $\theta_2$ and $\varphi_2$. Thus, if we have a stack of $N_f$ flavor branes, with $N_f\to\infty$,  we can distribute them in an homogeneous way along the directions transverse to the embeddings (\ref{embeddings}). As usual, the action for such a stack will be given by the  sum of a DBI and a WZ term, namely:
\beq
S_{flavor}\,=\,T_5\,\sum_{N_f}\,\Big[\,-\int_{{\cal M}_6}\,d^6\xi\,e^{{\phi\over 2}}\,
\sqrt{- \hat G_6} \,+\,\int_{{\cal M}_6} \hat C_6\,\Big]\,\,.
\label{Sflavor}
\eeq
The smearing procedure amounts to performing the following substitution on the DBI term of (\ref{Sflavor}):
\beq
-T_5\,\sum_{N_f}\,\,\int_{{\cal M}_6}\,d^6\xi\,e^{{\phi\over 2}}\,
\sqrt{- \hat G_6} \,\,\Longrightarrow\,\,
-{T_5 N_f\over (4\pi )^2}\,\,\int_{{\cal M}_{10}}\,
\,d^{10}\,x\,\,\sin\theta_1\,\sin\theta_2\,
e^{{\phi\over 2}}\,\,\sqrt{- \hat G_6}\,\,,
\eeq
where the factor $\sin\theta_1\,\sin\theta_2$ originates in the volume form of the space transverse to the embedding and $(4\pi)^2$ is a normalization factor that ensures that the total number of D5-branes is just $N_f$. Similarly, the WZ term of the system of smeared flavor branes is:
\beq
T_5\,\sum_{N_f}\,\int_{{\cal M}_6} \hat C_6\,\,\Longrightarrow\,\,
-{T_5\,N_f\over (4\pi )^2}\,\,\int_{{\cal M}_{10}}\,
{\rm Vol}(\,{\cal Y}_4^{1,2}\,)\,\wedge C_6\,\,,
\label{smearedWZ}
\eeq
and the minus sign is due to the different orientation of the worldvolume coordinates (\ref{wv-cordinates}) and those of the ten-dimensional space. In (\ref{smearedWZ}) 
${\rm Vol}(\,{\cal Y}_4^{1,2}\,)$ is the volume form of the four-dimensional space spanned by the directions $1$, $2$, $\hat 1$ and $\hat 2$, namely:
\beq
{\rm Vol}(\,{\cal Y}_4^{1,2}\,)\,=\,\sin\theta_1\,\sin\theta_2\,d\theta_1\,\wedge d\varphi_1\,
\wedge\, d\theta_2\,\wedge\, d\varphi_2\,=\,
\sigma^1\wedge \sigma^2\wedge \omega^1\wedge \omega^2\,\,.
\eeq
The cylinder embeddings just considered are extended along the $3$ and $\hat 3$ directions. However, there is nothing special in our background about these directions. Indeed, both in the metric and in the RR three-form, we are adopting a round ansatz which does not distinguish among the directions of the two three-spheres. Actually, by using an appropriate coordinate parameterization of the $\sigma^i$ and $\omega^i$ one-forms one can straightforwardly construct supersymmetric cylinder embeddings that span the $1, \hat 1$ or $2, \hat 2$ directions. The volume forms of the spaces transverse to these embeddings are clearly:
\beq
{\rm Vol}(\,{\cal Y}_4^{2,3}\,)\,=\,\sigma^2\wedge \sigma^3\wedge \omega^2\wedge \omega^3\,\,,
\qquad\qquad
{\rm Vol}(\,{\cal Y}_4^{1,3}\,)\,=\,\sigma^1\wedge \sigma^3\wedge \omega^1
\wedge \omega^3\,\,.
\eeq
To construct a backreacted supergravity solution with the same type of ansatz as in (\ref{ansatz}) we should consider a brane configuration that combines these three possible  types of  embeddings in an isotropic way.  The corresponding transverse volume form of this three-branch brane system would be:
\beq
{\rm Vol}(\,{\cal Y}_4\,)\,=\,{\rm Vol}(\,{\cal Y}_4^{1,2}\,)\,+\,{\rm Vol}(\,{\cal Y}_4^{2,3}\,)\,+\,{\rm Vol}(\,{\cal Y}_4^{1,3}\,)\,\,.
\label{totalVol}
\eeq
Notice that the WZ term of the action of the flavor branes can be written as:
\beq
S_{flavor}^{WZ}\,=\,T_5\,\int_{{\cal M}_{10}}\,\Omega\wedge C_6\,\,,
\eeq
where $\Omega$ is the following four-form:
\beq
\Omega\,=\,-{N_f\over 16\pi^2}\,\,{\rm Vol}(\,{\cal Y}_4\,)\,\,.
\eeq

Let us now write the DBI action for the D5-brane in terms of the smearing form $\Omega$. First of all, we notice that $\Omega$ is the sum of three decomposable pieces, namely:
\beq
\Omega\,=\,\Omega^{(1)}\,+\,\Omega^{(2)}\,+\,\Omega^{(3)}\,\,,
\eeq
where $\Omega^{(i)}$ is the transverse volume form  of the $i^{th}$ branch. Let us define the modulus of any of these components as:
\beq
\Big|\,\Omega^{(i)}\,\Big|\,=\,\sqrt{{1\over 4!}\,
\Omega^{(i)}_{M_1\cdots M_4}\,
\Omega^{(i)}_{N_1\cdots N_4}\,\,
\prod_{k=1}^{4}\,G^{M_k N_k}}\,\,.
\label{modulusOmega}
\eeq
In order to compute these moduli, it is convenient to express the $\Omega^{(i)}$'s in flat components with respect to the basis of one-forms (\ref{basis}):
\bear
&&\Omega^{(1)}\,=\,-
{N_f\over \pi^2}\,\,e^{-4f-2h-2g}\,\,\,e^1\wedge e^2\wedge e^{\hat 1}
\wedge e^{\hat 2}\,\,,\rc\rc
&&\Omega^{(2)}\,=\,-
{N_f\over\pi^2}\,\,e^{-4f-2h-2g}\,\,\,e^1\wedge e^3\wedge e^{\hat 1}
\wedge e^{\hat 3}\,\,,\rc\rc
&&\Omega^{(3)}\,=\,-
{N_f\over \pi^2}\,\,e^{-4f-2h-2g}\,\,\,e^2\wedge e^3\wedge e^{\hat 2}
\wedge e^{\hat 3}\,\,.
\eear
It follows from the previous expressions that:
\beq
\Big|\,\Omega^{(1)}\,\Big|\,=\,\Big|\,\Omega^{(2)}\,\Big|\,=\,
\Big|\,\Omega^{(3)}\,\Big|\,=\,{N_f\over \pi^2}\,\,e^{-4f-2h-2g}\,\,.
\eeq
The DBI action for the first branch in the standard coordinate system can be written in terms of $|\,\Omega^{(1)}\,|$. Indeed, one can prove that this action is given by:
\beq
-{T_5 N_f\over (4\pi )^2}\,\,\int_{{\cal M}_{10}}\,
\,d^{10}\,x\,\,\sin\theta_1\,\sin\theta_2\,
e^{{\phi\over 2}}\,\,\sqrt{- \hat G_6} \,=\,
-T_5\,\int_{{\cal M}_{10}}\,d^{10}x\,\,\,e^{{\phi\over 2}}\,\,
\sqrt{-G}\,\,\Big|\,\Omega^{(1)}\,\Big|\,\,.
\eeq
It is now clear how to generalize this result to include the three branches, namely:
\beq
S_{flavor}^{DBI}\,=\,-T_5\,\int_{{\cal M}_{10}}\,d^{10}x\,\,\,e^{{\phi\over 2}}\,\,
\sqrt{-G}\,\,\sum_i\,\Big|\,\Omega^{(i)}\,\Big|\,\,.
\eeq
Thus the total action of the brane distribution can be written in terms of the four-form
$\Omega$.

\subsection{The backreaction}

Let us consider the coupled gravity plus branes system with action:
\beq
S\,=\,S_{IIB}\,+\,S_{flavor}\,\,,
\label{totalS}
\eeq
where $S_{IIB}$ is the action, in the Einstein frame, of type IIB supergravity for the metric, dilaton and RR three-form $F_3$:
\beq
S_{IIB}\,=\,{1\over 2\kappa_{10}^2}\,\,
\int d^{10} x\,\sqrt{-G}\,\Big[\,R\,-\,{1\over 2}\,\partial_M \phi\,\partial^M\phi\,-\,
{1\over 12}\,e^{\phi}\,F_3^2\,\Big]\,\,,
\eeq
and $S_{flavor}$ is the action for a set of smeared  flavor D5-branes, given by:
\beq
S_{flavor}\,=\,\,-T_5\,\int_{{\cal M}_{10}}\,d^{10}x\,\,\,e^{{\phi\over 2}}\,\,
\sqrt{-G}\,\,\sum_i\,\Big|\,\Omega^{(i)}\,\Big|\,+\,
T_5\,\int_{{\cal M}_{10}}\,\Omega\wedge C_6\,\,.
\label{total-smeared-action}
\eeq
In (\ref{total-smeared-action}) $\Omega$ is the four-form that encodes the RR charge distribution of the smeared stack of D5-branes, while the moduli $\big|\,\Omega^{(i)}\,\big|$ of its decomposable parts determine the mass distribution of the stack.  In order to determine how the smeared action (\ref{total-smeared-action}) for the flavor branes affects the equations of motion of the RR forms, it is convenient to recall that, in the Einstein frame, the field strength $F_7=dC_6$ is related to $F_3$ as 
$F_7=-e^{\phi}\,^*F_3$. Then, the equation of motion of $C_6$ derived from the action (\ref{totalS}) is just:
\beq
d\Big(\,e^{-\phi} \,{}^*F_7\,\Big)\,=\,-2\,T_5\,\kappa_{10}^2\,\Omega\,\,.
\eeq
Using the fact that, in our conventions:
\beq
T_5\,=\,{1\over (2\pi)^5}\,\,,\qquad\qquad
2\kappa_{10}^2\,=\,(2\pi)^7\,\,,
\eeq
we can rewrite the equation for $C_6$ as:
\beq
d\Big(\,e^{-\phi} \,{}^*F_7\,\Big)\,=\,-4\pi^2\,\Omega\,\,.
\eeq
Since $e^{\phi}\,F_3=-{}^*F_7$, this  equation is equivalent to the following violation
of the Bianchi identity of $F_3$:
\beq
dF_3\,=\,4\pi^2\,\Omega\,=\,-{N_f\over 4}\,{\rm Vol }({\cal Y}_4)\,\,,
\label{Bianchi-violation}
\eeq
where ${\rm Vol }({\cal Y}_4)$ has been written in (\ref{totalVol}). It is clear from this last equation that, in order to find a solution including the backreaction of the smeared flavor branes, we must modify our ansatz for the RR three-form $F_3$. Actually, we shall try to find a solution in which 
\beq
F_3\,=\,{\cal F}_3\,+\,f_3\,\,,
\label{F3f3}
\eeq
where ${\cal F}_3$ is just given by the same ansatz (\ref{F3ansatz}) as in the unflavored case (with $d{\cal F}_3=0$) and $f_3$ is a new term that gives rise to the violation 
(\ref{Bianchi-violation}) of the Bianchi identity. It is easy to see that one can take:
\begin{equation}
f_3=\frac{N_f}{8}\epsilon_{ijk}(\omega^i-\frac{\sigma^i}{2})\wedge\sigma^j\wedge\sigma^k\,.
\label{f3}
\end{equation}
By plugging the modified ansatz (\ref{F3f3})-(\ref{f3}) for the RR three-form into the supersymmetric variations of the dilatino and gravitino of type IIB supergravity one obtains a system of first-order BPS equations for the different functions of the ansatz. The corresponding calculations are presented in appendix \ref{BPSapp}. In appendix 
\ref{EOMapp} we check that any solution of the BPS equations solves the equations of motion.

Let us point out that, as happened for the unflavored case, one can consistently truncate the BPS equations by taking $w=\gamma=\kappa=0$. In some cases this truncation represents the UV limit of the solutions of the  full BPS system of equations.  In the next section we will study in detail these simplified solutions and we will get some  interesting information about the corresponding gauge theory duals. The analysis of the complete BPS equations will be performed in section \ref{untruncated-flavor}.

\setcounter{equation}{0}
\section{The truncated system with flavor}
\label{truncated-flavor}

In this section we will analyze the truncation of the general system of BPS equations that corresponds to taking $w=\gamma=\alpha= V=0$. In this case the equations of appendix  \ref{BPSapp} for the dilaton and for the remaining functions $h$ and $g$ of the metric reduce to:
\bear
&&\phi'\,=\,N_c\,e^{-3g}\,\,-\,{3\over 4}\,(\,N_c-4N_f\,)\,e^{-g-2h}\,\,,\rc\rc
&&h'\,=\,{1\over 2}\,e^{g-2h}\,+\,{N_c-4N_f\over 2}\,\,e^{-g-2h}\,\,,\rc\rc
&&g'\,=\,e^{-g}\,-\,{1\over 4}\,e^{g-2h}\,-\,N_c\,e^{-3g}\,+\,{N_c-4N_f \over 4}\,
e^{-g-2h}\,\,.
\label{ab-system-flavor}
\eear

By inspecting the system (\ref{ab-system-flavor}) one readily realizes that there are some special solutions for which the metric functions $h$ and $g$ are constant. Actually these solutions only exist when $N_c<2N_f$ and the expressions of $g$ and $h$ are the following:
\beq
e^{2g}\,=\,4N_f-N_c\,\,,
\qquad\qquad
e^{2h}\,=\,{1\over 4}\,\,{(4N_f-N_c)^2\over 2N_f-N_c}\,\,,
\qquad\qquad (\,N_c\, <\,2N_f\,)\,\,,
\eeq
while the dilaton grows linearly with the holographic coordinate $r$, namely:
\beq
\phi\,=\,{2(3N_f-N_c)\over \big[\,4N_f-N_c]^{{3\over 2}}}\,\,r\,\,+\,\,\phi_0\,\,.
\eeq

Let us next consider solutions for which the function $h$ is not constant. In this case we can use $\rho=e^{2h}$ as radial variable as in (\ref{rho}) and one can define the function 
$F(\rho)$ as in (\ref{F}). The BPS equation for $F(\rho)$ is now:
\beq
{dF\over d\rho}\,=\,{(F-N_c)\,\Big(\,2-{F\over 2\rho}\,\Big)\,-\,{2N_f\over \rho}\,\,F
\over F+N_c-4N_f}\,\,,
\label{F-ab-flavor}
\eeq
while the equation for the dilaton  as a function of $\rho$ can be written as:
\beq
{d\phi\over d\rho}\,=\,{N_c\over F(F+N_c-4N_f)}\,\Big[\,1\,-\,{3\over 4\rho}\,
\Big(\,1\,-\,{4N_f\over N_c}\,\Big)\,F\,\Big]\,\,.
\label{dilaton-ab-flavor}
\eeq
Moreover,  from the second equation in (\ref{ab-system-flavor}) we can obtain the relation between the two radial variables $r$ and $\rho$, namely:
\beq
{dr\over d\rho}\,=\,{\sqrt{F(\rho)}\over F(\rho)+N_c-4N_f}\,\,.
\label{r-rho-flavored-ab}
\eeq
Notice that, contrary to the unflavored case (see eq. (\ref{r-rho-unflavored})), the sign of the right-hand side of (\ref{r-rho-flavored-ab}) could be negative when $N_f\not=0$. This means that we have to be careful in identifying the UV and IR domains in terms of the new radial variable $\rho$.  Let now study different solutions of eqs. (\ref{F-ab-flavor})-(\ref{dilaton-ab-flavor}).

\subsection{Linear dilaton backgrounds}
\label{FMnas-abe-subsection}

It is clear from (\ref{F-ab-flavor}) that in this case $F=N_c$ is no longer a solution of the equations. However, there are solutions for which this constant value of $F$ is reached asymptotically when $\rho\to\infty$. Indeed, one can check this fact by solving (\ref{F-ab-flavor})  as an expansion in powers  of $1/\rho$. One gets:
\beq
F\,=\,N_c\,+\, N_c\, N_f\,{1\over \rho}\,-\,
{3\over 4}\, N_c\, N_f\,(N_c-4N_f)\,{1\over \rho^2}\,+\,\cdots\,\,,
\qquad\qquad (\rho\to\infty)\,\,.
\label{UVF}
\eeq
By plugging the expansion (\ref{UVF}) into (\ref{dilaton-ab-flavor}) one can prove that, when $N_c\not=2N_f$,  these solutions have a  dilaton  that depends  linearly on $\rho$  in the UV and, actually,  one can verify that:
\beq
{d\phi\over d\rho}\,=\,{1\over 2(N_c-2N_f)}\,-\,
{3N_c^2\,-\,12 N_c N_f\,+\,16N_f^2\over 8(N_c-2N_f))^2}\,\,\,{1\over \rho}
\,+\,\cdots\,\,,
\qquad\qquad (\rho\to\infty)\,\,.
\label{UVdilaton}
\eeq
Notice the different large $\rho$  behavior of the dilaton in the two cases $N_c>2N_f$ and  $N_c<2N_f$. Indeed, when $N_c>2N_f$ the dilaton grows linearly with the holographic coordinate $\rho$ (the behavior expected for a confining theory), while for $N_c<2N_f$ the field $\phi$ decreases linearly with $\rho$. This seems to suggest that the beta function of the dual gauge theory depends on $N_c$ and $N_f$ through the combination $N_c-2N_f$.  Actually, one can verify that when $N_c>2N_f$ the sign of $dr/d\rho$ is positive, while if $N_c<2N_f$    the derivative $dr/d\rho$ changes its sign and $r$ decreases when $\rho$ increases.  Indeed, by plugging the expansion (\ref{UVF}) on the right-hand side of (\ref{r-rho-flavored-ab}) one gets:
\beq
{dr\over d\rho}\,=\,{\sqrt{N_c}\over 2(N_c-2N_f)}\,+\,{\sqrt{N_c}\,N_f^2\over 2(N_c-2N_f)^2}\,{1\over \rho}\,+\,\cdots\,\,,
\qquad\qquad (\rho\to\infty)\,\,.
\label{r-rho-aympt}
\eeq
\begin{figure}[ht]
\begin{center}
\includegraphics[width=0.95\textwidth]{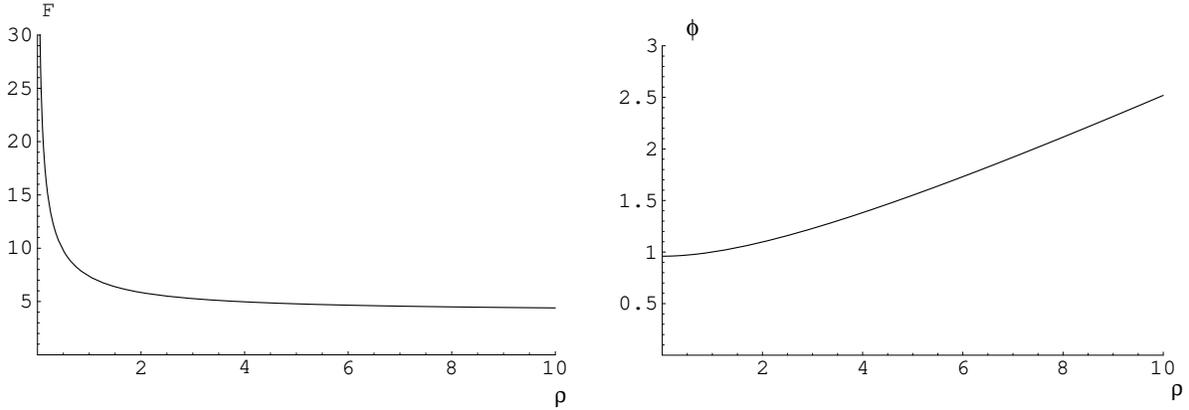}
\end{center}
\caption[UVFPhi]{The function $F$ and the dilaton for the case $N_c=4$, $N_f=1$.} 
\label{FPhiNc4Nf1}
\end{figure}

\begin{figure}[ht]
\begin{center}
\includegraphics[width=0.35\textwidth]{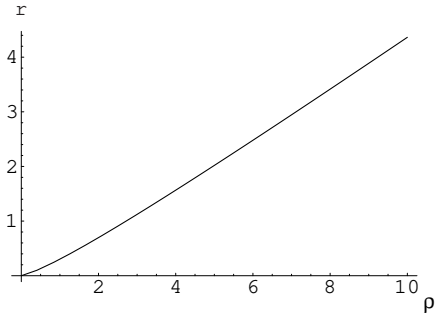}
\end{center}
\caption[r-rho]{Relation between the radial variables $r$ and $\rho$  for the case $N_c=4$, $N_f=1$.}
\label{r-rhoNf4Nc2}
\end{figure}

\begin{figure}[ht]
\begin{center}
\includegraphics[width=0.95\textwidth]{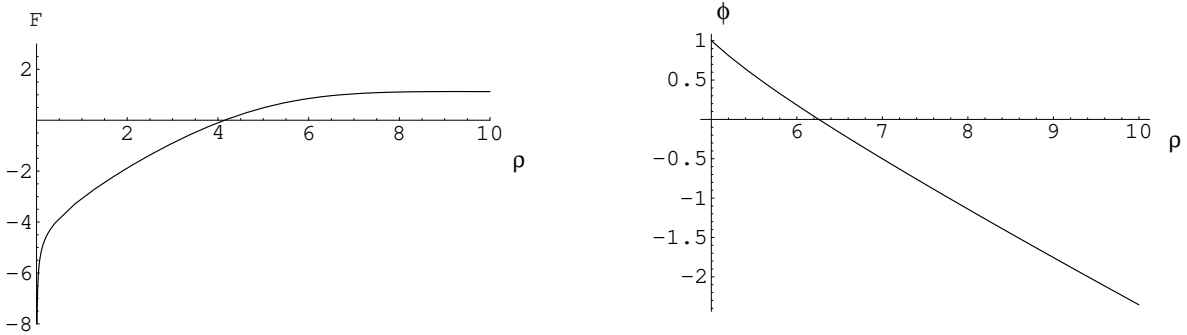}
\end{center}
\caption[UVPhiNc1Nf1]{The function $F$ and the dilaton $\phi$ for the case $N_c=1$, $N_f=1$.}
\label{FPhiNc1Nf1}
\end{figure}

\begin{figure}[ht]
\begin{center}
\includegraphics[width=0.35\textwidth]{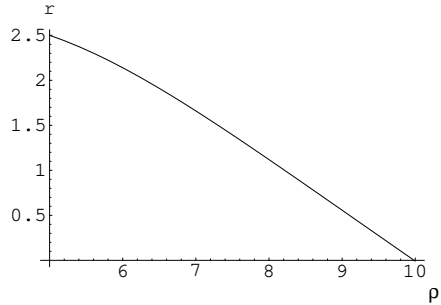}
\end{center}
\caption[UVPhiNc1Nf1]{The function $r(\rho)$ for the case $N_c=1$, $N_f=1$.
This curve is obtained by integrating eq. (\ref{r-rho-flavored-ab}) in the region in which $F$ is non-negative.}
\label{r-rhoNc1Nf1}
\end{figure}

The first term on the right-hand side of (\ref{r-rho-aympt}) is clearly dominant for $\rho\to\infty$. Its sign is the same  as the one in the combination $N_c-2N_f$, which shows that, at least in the $\rho\to\infty$ region, the relation between the two radial variables $r$ and $\rho$ is the one described above. One can confirm this behavior by numerical integration of the differential equations (\ref{F-ab-flavor}), (\ref{dilaton-ab-flavor}) and (\ref{r-rho-flavored-ab}). In figures \ref{FPhiNc4Nf1} and \ref{r-rhoNf4Nc2} we have plotted the result of this integration for a case in which $N_c>2N_f$, namely $N_c=4$, $N_f=1$.  We have   integrated (\ref{F-ab-flavor}) by imposing the behavior (\ref{UVF}) on $F(\rho)$  for large $\rho$. Once $F(\rho)$ is known one can obtain $\phi(\rho)$ and 
$r(\rho)$ by direct integration of the right-hand sides of eqs. (\ref{dilaton-ab-flavor}) and (\ref{r-rho-flavored-ab}). We notice in figure \ref{FPhiNc4Nf1} that $F$ diverges for $\rho\to 0$, while the dilaton $\phi$ remains finite for small $\rho$. It is easy to characterize analytically these behaviors. Indeed, for small $\rho$ it is also possible to solve the equations (\ref{F-ab-flavor}) and (\ref{dilaton-ab-flavor}) in a series expansion near $\rho\approx 0$. For the function $F(\rho)$ one has:
\beq
F(\rho)\,=\,{c_0\over \sqrt{\rho}}\,+\,2(N_c\,-\,4N_f)\,-\,
{(N_c-4N_f)^2\over c_0}\,\,\sqrt{\rho}\,+\,
\Big(\,{4\over 3}\,+\,{2(N_c-4N_f)^3\over c_0^2}\,\Big)\,\rho\,+\,\cdots\,\,,
\label{F-truncated-IR-flavor}
\eeq
where $c_0$ is an integration constant which, for consistency, must be taken to be positive. In general, only for one particular value of $c_0\not=0$ does one get a solution that behaves as in (\ref{UVF}) for $\rho\to\infty$. Similarly, the dilaton for $\rho\approx 0$ behaves as:
\beq
{d\phi\over d\rho}\,=\,-{3(N_c-4N_f)\over 4 c_0}\,\,{1\over \sqrt{\rho}}\,+\,
{9(N_c-4N_f)^2\over 4 c_0^2}\,+\,\cdots\,\,,
\qquad
(\rho\approx 0)\,\,.
\label{Phi-truncated-IR-flavor}
\eeq
(Compare eqs. (\ref{F-truncated-IR-flavor}) and (\ref{Phi-truncated-IR-flavor}) with the ones corresponding to the unflavored solutions, namely (\ref{F-vs-rho-IR})  and (\ref{Phi-vs-rho-IR})).
Notice that, as in our numerical integration,  eq. (\ref{Phi-truncated-IR-flavor})  implies that $\phi$ is regular at $\rho\approx 0$ when $c_0\not=0$ (although ${d\phi\over d\rho}$ is divergent). One can also  easily  get the value of the derivative $dr/d\rho$ for small values of $\rho$, which is given by:
\beq
{dr\over d\rho}\,=\,{\rho^{{1\over 4}}\over \sqrt{c_0}}\,-\,
{2(N_c-4N_f)\over c_0^{{3\over 2}}}\,\rho^{{3\over 4}}\,
+\,\cdots\,\,,
\qquad
(\rho\approx 0)\,\,.
\eeq
From the plot of $r$ versus $\rho$ of figure \ref{r-rhoNf4Nc2} we notice that $r$ grows monotonically with $\rho$ in this $N_c>2N_f$ case.  This means that the UV region $r\to\infty$ corresponds to large values of $\rho$.  Below we will study the beta function of the gauge theory and we will conclude that the theory is asymptotically free when $N_c>2N_f$, while it develops a Landau pole in the UV when  $N_c<2N_f$. We can confirm this statement by looking at the result of the numerical integration when $N_c<2N_f$. 
In figure \ref{FPhiNc1Nf1} we present the result of this integration for $N_c=1$ and $N_f=1$. As before, we impose the behavior (\ref{UVF}) for large values of  $\rho$. We notice that $F(\rho)$ becomes negative at some finite value $\rho_*$ of the coordinate $\rho$, which means that the space ends at $\rho=\rho_*$ and we should consider the region $\rho\ge \rho_*$ as the one that is physically sensible. Actually, in this region the dilaton $\phi$ decreases with $\rho$. A glance at the $r-\rho$ relation displayed in figure 
\ref{r-rhoNc1Nf1} shows that $r$ decreases with $\rho$ and, actually, large values of $\rho$ correspond to  small values of $r$, \ie\ to the IR region of the dual gauge theory. It is also clear from figure \ref{r-rhoNc1Nf1}  that there is a maximal value of $r$, which corresponds to the minimal value $\rho_*$ of $\rho$. This fact is signaling the presence of a Landau pole in the UV of the gauge theory dual. 

When $N_c=2N_f$ the expansion (\ref{UVdilaton}) is clearly not valid and we are in a borderline case. One can prove that in this case
\beq
{d\phi\over d\rho}\,=\,{2\over N_c^2}\,\,\rho\,-\,{1\over N_c}\,+\,\cdots\,\,,
\qquad\qquad
(N_c=2N_f, \,\rho\to\infty)\,\,,
\eeq
and, thus, $\phi$ grows quadratically with $\rho$ when $\rho\to\infty$. Similarly, when $N_c=2N_f$, the relation between the two radial variables $r$ and $\rho$  is:

\beq
{dr\over d\rho}\,=\,{2\over (N_c)^{{3\over 2}}}\,\,\rho\,+\,{2\over \sqrt{N_c}}\,+\,\cdots\,\,,
\qquad\qquad
(N_c=2N_f, \,\rho\to\infty)\,\,,
\label{r-rho-border}
\eeq
which means that $r\sim \rho^2$.  By combining the last two equations we conclude that
the dilaton grows linearly with the $r$ coordinate also in this $N_c=2N_f$ case.

\subsection{Flavored $G_2$ cone}
\label{G2cones-truncated-subsection}

Let us now consider the solution of the equations (\ref{F-ab-flavor}) and (\ref{dilaton-ab-flavor}) that leads to a metric which is asymptotically a $G_2$-cone with constant dilaton  in the UV.  It can be checked that there exists a solution of (\ref{F-ab-flavor}) which can be expanded for large values of $\rho$ as:
\beq
F\,=\,{4\over 3}\,\,\rho\,+\,4(N_f-N_c)\,+\,
{15 N_c^2-39N_c N_f\,+\,24 N_f^2\over \rho}\,+\,\cdots\,\,.
\label{G2-F-flavored}
\eeq
The corresponding expansion for $d\phi/d\rho$ is:
\beq
{d\phi\over d\rho}\,=\,{9N_f\over 4}\,{1\over \rho^2}\,+\,{27\over 16}\,N_c\,(N_c+2N_f)\,
{1\over \rho^3}\,+\,\cdots\,\,,
\label{G2-phi-flavored}
\eeq
which can be integrated as:
\beq
\phi\,=\,\phi_{\infty}\,-\,{9N_f\over 4}\,{1\over \rho}\,-\,{27\over 32}\,N_c\,(N_c+2N_f)\,
{1\over \rho^2}\,+\,\cdots\,\,.
\label{G2-int-phi-flavored}
\eeq
Notice that, when $N_f=0$, the expansions (\ref{G2-F-flavored}) and (\ref{G2-phi-flavored})  reduce to the ones displayed in eqs. 
(\ref{G2-F-unflavored}) and (\ref{G2-phi-unflavored}). To find the solution in the whole range of the radial coordinate one can integrate numerically the system (\ref{F-ab-flavor})-(\ref{dilaton-ab-flavor}) by imposing the asymptotic behavior (\ref{G2-F-flavored}) to the function $F(\rho)$. The results for $N_c\ge 2N_f$ are similar to the ones found in subsection \ref{truncated-unflavor-section} for the unflavored system and, in particular, the solution is well-defined for all possible values of the coordinate $\rho$.  On the contrary, when $N_c<2N_f$  the solution only makes sense when $\rho$ is greater than some $\rho_0$, with $\rho_0>0$.  To illustrate this fact let us consider a particular case with $N_c<2N_f$, namely $N_c=N_f$. In this case the BPS system (\ref{F-ab-flavor})-(\ref{dilaton-ab-flavor}) can be integrated analytically. We first notice that  the subleading terms in  (\ref{G2-F-flavored}) cancel when $N_c=N_f$. Actually, one can check that in this case the leading term in (\ref{G2-F-flavored}) is an exact solution of the differential eq. (\ref{F-ab-flavor}), namely:
\beq
F\,=\,\,{4\over 3}\,\,\rho\,\,,\qquad\qquad (N_f\,=\,N_c)\,\,.
\eeq
Plugging this result into the equation (\ref{dilaton-ab-flavor}) for $\phi$, one gets:
\beq
\phi\,=\,\phi_{\infty}\,+\,\log\big(\,1\,-\,{9N_c\over 4\rho}\,\big)
\,\,,\qquad\qquad (N_f\,=\,N_c)\,\,.
\label{G2-phi-exact}
\eeq 
As a check  one can verify that the expansion of (\ref{G2-phi-exact}) for large values of $\rho$ coincides with the one written in (\ref{G2-int-phi-flavored}) for $N_c=N_f$.  Notice that the dilaton  in (\ref{G2-phi-exact}) diverges when $\rho=\rho_*$, where $\rho_*$ is given by:
\beq
\rho_*\,=\,{9N_c\over 4}\,\,.
\eeq
To understand the origin of this divergence it is interesting to look at the change of radial variables in this case. Actually, when $N_c=N_f$ the equation (\ref{r-rho-flavored-ab}) that determines the function $r(\rho)$ can be integrated to give:
\beq
r(\rho)\,=\,r_0\,+\,\sqrt{3\rho}\,+\,{3\sqrt{3N_c}\over 4}\,\,
\log\Big[\,
{2\sqrt{\rho}-3\sqrt{N_c}\over 2\sqrt{\rho}+3\sqrt{N_c}}\,\Big]
\,\,,\qquad\qquad (N_f\,=\,N_c)\,\,.
\label{r-rho-exact}
\eeq
It is clear from this expression that $r(\rho)\to-\infty$ if $\rho\to\rho_*$. Thus, as $r$ must be non-negative, we should restrict $\rho$ to the range $\rho_0\le \rho<+\infty$, where 
$\rho_0>\rho_*$ is determined by the condition $r(\rho_0)=0$ (the actual value of 
$\rho_0$ depends on the value chosen for the integration constant $r_0$ in (\ref{r-rho-exact})).

\section{The general system with flavor ($N_c\ge 2N_f$)}
\label{untruncated-flavor}

Let us now consider the BPS equations of appendix \ref{BPSapp} in full generality. In this section we will restrict ourselves to the case $N_c\ge 2N_f$ which, according to our analysis of the truncated system in section \ref{truncated-flavor}, is expected to give more sensible solutions describing an asymptotically free gauge theory.  First of all, as in subsection \ref{untruncated-unflavored-subsection}, we are going to  write these equations in terms of the variable $\rho=e^{2h}$. The equation
for $F=e^{2g}$ as a function of $\rho$ can be written as in (\ref{dF}), where now the coefficients $A$, $\tilde A$, $D$ and $\tilde D$ are given by:
\bear
&&A\,=\,\Big[\,2\,-\,(1-w^2)\,{F\over 2\rho}\,\Big]\,\Big[\,F\,-\,N_c\,\Big]\,
-2N_f\,{F\over \rho}\,+\,
N_c\,w\,(w-\gamma)\,{F\over \rho}\,\,,\rc\rc
&&\tilde A\,=\,2N_c\,(w-\gamma)\,\sqrt{{F\over \rho}}\,\,,\rc\rc
&&D\,=\,(1-w^2)\,(F+N_c)\,-\,4N_f\,+\,
2N_c\,w\,(w-\gamma)\,\,,\rc\rc
&&\tilde D\,=\,{N_c\over 4}\, V\,\sqrt{{F\over \rho}}\,+\,N_c\,
(w-\gamma)\,\sqrt{{\rho\over F}}\,-\,2w\sqrt{F\rho}\,\,,
\label{A-D-flavor}
\eear
with $V$ being the function of $w$, $\gamma$ and the constant $\kappa$ written in 
(\ref{V-general}). In this case we can also represent $\beta$ and $\tilde\beta$ as in (\ref{betas-Lambdas}), but  now the functions $ \Lambda$ and $\tilde \Lambda$ are the following:
\bear
&&\Lambda\,=\,\rho\,+\,{1-w^2\over 4}\,F\,+\,{N_c\over 4}\,
(1-\,{4N_f\over N_c}\,+\,w^2-2 w\gamma)\,-\,
{N_c\rho\over 3F}\,\,,\rc\rc
&&\tilde\Lambda\,=\,{N_c\over 24}\, V\,\sqrt{F\over \rho}\,-\,w\sqrt{\rho F}\,+\,
{N_c\over 2}\,(w-\gamma)\,\, \sqrt{{\rho\over F}}\,\,.
\label{Lambdas-flavor}
\eear
Similarly, the equations that govern $w$ and $\gamma$ can be written as in 
(\ref{w-gamma-rho}), with the coefficients $B$, $\tilde B$, $C$ and $\tilde C$ given by:
\bear
&&B\,=\,{2N_c\over 3}\,
\Bigg[\,{ V\over 4\rho}\,-\,3\,{w-\gamma\over F}\,\Bigg]\,\,,\rc\rc
&&\tilde B\,=\,{4\over 3}\,\Bigg[\,\Big(\,3\,-\,{2N_c\over F}\,\Big)\,\sqrt{{\rho\over F}}\,-\,
{3\over 4}(1-w^2)\,\sqrt{{F\over \rho}}\,\Bigg]\,\,,\rc\rc
&&C\,=\,{2\over 3}\,\Bigg[\,{ V\over 4\rho}\,F\,+\,3(w-\gamma)\,\Bigg]\,\,,\rc\rc
&&\tilde C\,=\,{4\over 3}\,\Bigg[\,\sqrt{{\rho\over F}}\,-\,{3\over 4}\,\,
\Big(1-{4N_f\over N_c}\,+\,w^2-2w\gamma\,\Big)\,\sqrt{{F\over \rho}}\,\Bigg]\,\,.
\eear

Finally, the BPS equation for the dilaton can be represented as in (\ref{dilaton-in-rho}),
where now the functions $E$ and $\tilde E$ are:
\bear
&&E\,=\,N_c\,\Big[\,{1\over F}\,-\,{3\over 4\rho}\,
\Big(\,1-{4N_f\over N_c}\,+\,w^2-2w\gamma\Big)
\Big]\,\,,\rc\rc
&&\tilde E\,=\,-N_c\,\Big[\,{ V\over 8\rho}\,\,\sqrt{F\over \rho}\,+\,
{3(w-\gamma)\over 2\sqrt{F\rho}}\,\Big]\,\,.
\eear

As in the unflavored case, in order to solve the BPS system we have to fix the initial conditions of the different functions of the ansatz, as well as the constant $\kappa$. To determine these values we follow, step by step, the procedure employed in subsection \ref{untruncated-unflavored-subsection} for the unflavored case. Namely, we will impose certain regularity conditions in the IR.  Notice that, according to our analysis of the truncated system in section \ref{truncated-flavor},  we expect that for $N_c\ge 2 N_f$ this IR region will correspond to  $\rho\approx 0$.

First of all, let us point out that the arguments given in order to arrive at eq. (\ref{initialw}) are still valid in this case and, therefore, we will continue to require that $w(\rho=0)=1$. Moreover, the requirement that $F(\rho)$ is regular at $\rho=0$ is also quite natural and, thus, we will also assume that (\ref{F-IR}) holds in this flavored case. Notice that the cycle $\Sigma$ defined in (\ref{mixc}) also collapses in the IR in the present case (the induced metric on $\Sigma$ is still given by (\ref{ds-on-Sigma})) and, therefore, we should impose the vanishing of the corresponding  pullback of $F_3$. Actually, 
it is immediate from (\ref{f3}) that the pullback on $\Sigma$ of the flavor contribution $f_3$ to the RR three-form is:
\beq
{f_3}{\big|_\Sigma}\,=\,{3N_f\over 8}\,\,\sigma^1\wedge\sigma^2\wedge \sigma^3\,\,,
\eeq
and thus, the total pullback of $F_3$ is:
\beq
{F_3}{\big|_\Sigma}\,=\,{N_c\over 4}\,\,
\Big[\,\kappa\,-\,{1\over 2}\,+\,{3N_f\over 2 N_c}\,\Big]
\,\,\sigma^1\wedge\sigma^2\wedge \sigma^3\,\,.
\eeq
Thus, we will require that $\kappa$ takes the value:
\beq
\kappa\,=\,{1\over 2}\,-\,{3N_f\over 2 N_c}\,\,.
\label{kappa-flavors}
\eeq
The next requirement that we will implement is the regularity of the dilaton in the IR. Since the reasoning that leads to the behavior (\ref{divergent-dilaton}) is also valid for the flavored system, we conclude that we should also impose (\ref{V0=0}) in the present case. Moreover, by substituting $w=1$ in the expression for $V$ in (\ref{V-general}), we get:
\beq
V_0\,=\,8\Big(\,\kappa-{1\over 2}+{3N_f\over 2 N_c}\,
\Big)\,\,,
\eeq
which vanishes precisely when $\kappa$ is given by the value displayed in (\ref{kappa-flavors}). As we discussed in subsection \ref{untruncated-unflavored-subsection}, the vanishing of $V_0$ is not sufficient to ensure the finiteness of $\phi$ at the origin. Indeed, in addition, we should require (\ref{Lambda=0}), \ie\ that $\Lambda$ is also vanishing at $\rho=0$. From the expression of $\Lambda$ given in (\ref{Lambdas-flavor}) one discovers that this condition determines the IR value of $\gamma$ to be:
\beq
\gamma(\rho=0)\,=\,1\,-\,{2N_f\over N_c}\,\,.
\label{gamma-initial-flavored}
\eeq
Notice that the only freedom left by our IR regularity conditions is the value $F_0$ of the function $F(\rho)$ at $\rho=0$.  By changing this value of $F_0$ we can select some particular classes of solutions. We are mostly interested in the  backgrounds for which the dilaton grows linearly with the holographic coordinate in the UV and such that the function $F(\rho)$ reaches a constant value when $\rho\to\infty$. Those backgrounds are the flavored analogue of the ones studied in \cite{Maldacena:2001pb} and can be naturally interpreted as the gravity dual of 2+1 dimensional gauge theories with quarks transforming in the fundamental representation of the gauge group.  They will be obtained in subsection \ref{FlaMnas-subsection} by fine tuning $F_0$ to some particular value, that depends on the numbers of colors and flavors. In the UV region $\rho\to\infty$ we expect that these new backgrounds will coincide with the solutions of the truncated system studied in subsection \ref{FMnas-abe-subsection}, while for $\rho\to 0$ a significant difference between the truncated and untruncated solutions is expected.

Notice that the initial conditions (\ref{initialw}) and (\ref{gamma-initial-flavored}) ensure that the $(\rho, \sigma^i)$ part of the metric is of the form (\ref{rho-sigma-metric}). Nevertheless, in this flavored case the explicit calculation of the scalar curvature  for the linear dilaton solutions shows that the metric is singular at the origin of the radial coordinate. However, as our initial conditions are such that the dilaton is finite at the origin, the value of the $g_{tt}$ component of the metric is also bounded and then, according to the criterium of  \cite{Maldacena:2000mw}, the singularity is ``good" and the background can be used to extract non-perturbative information of the dual gauge theory.

As in subsection \ref{G2cones-truncated-subsection} we will also have backgrounds such that their  metric asymptotes in the UV to  that of a $G_2$ cone  with constant dilaton.  They will be briefly discussed in subsection \ref{G2cones-untruncated-subsection}.

\subsection{Asymptotic linear dilaton}
\label{FlaMnas-subsection}

As explained above,
we are interested in solutions of the BPS equations such that asymptotically $F$ is constant. Actually, by solving the BPS system in powers of $1/\rho$,
one can check that there are  solutions in which $F$ has the following  asymptotic 
behavior:
\beq
F\,=\,N_c\,+\,{a_1\over \rho}\,+\,{a_2\over \rho^2}\,+\,{a_3\over \rho^3}\,+\,\cdots\,\,,
\label{F-inseries-untruncated}
\eeq
where the coefficients $a_1$, $a_2$ and $a_3$ are given by:
\bear
&&a_1\,=\,N_f\, N_c\,\,,\rc\rc
&&a_2\,=\,-{3\over 4}\, N_c N_f(\,N_c\,-\,4N_f\,)\,\,,\rc\rc
&&a_3\,=\,{N_f\,N_c\over 16}\,
\Big[\,21N_c^2\,-\,148\,N_f\, N_c\,+\,240\,N_f^2\,\Big]\,\,.
\label{a-coefficients-untruncated}
\eear
Notice that the first two terms in (\ref{F-inseries-untruncated}) and 
(\ref{a-coefficients-untruncated}) coincide with the one written in (\ref{UVF}) for the truncated system.  Similarly, the functions $w$ and $\gamma$ can be represented as:
\bear
&&w\,=\,{b_1\over \rho}\,+\,{b_2\over \rho^2}\,+\,{b_3\over \rho^3}\,+\,\cdots\,\,,\rc\rc
&&\gamma\,=\,{c_1\over \rho}\,+\,{c_2\over \rho^2}\,+\,{c_3\over \rho^3}\,+\,\cdots\,\,,
\label{wgamma-inseries-untruncated}
\eear
where the coefficients $b_i$ and $c_i$ are the following:
\bear
&&b_1\,=\,c_1\,=\,{1\over 2}\,\,(N_c\,-3N_f)\,\,,\rc\rc
&&b_2\,=\,c_2\,=\,{5\over 8}\,\,(N_c\,-3N_f)\,(N_c-2N_f)\,\,,\rc\rc
&&b_3\,=\,{1\over 32}\,\,(N_c\,-3N_f)\,
\Big[\,49 N_c^2\,-\,184\,  N_c\,N_f\,+\,204\,N_f^2\,\Big]\,\,,\rc\rc
&&c_3\,=\,{1\over 32}\,\,(N_c\,-3N_f)\,
\Big[\,49 N_c^2\,-\,208\,N_f N_c\,+\,252\,N_f^2\,\Big]\,\,.
\label{bc-coefficients-untruncated}
\eear
By plugging the above series for $F$, $w$ and $\gamma$ in the equation for $\phi$,
one can also get the UV behavior of the dilaton as a power series in $1/\rho$. Actually, the first two terms in this expansion are just the ones written in (\ref{UVdilaton}). For $N_c>2N_f$  this means that, asymptotically, the dilaton grows linearly with the holographic coordinate $\rho$ as:
\beq
\phi\sim {\rho\over 2(N_c-2N_f)}\,+\,o(\log \rho)\,\,,\qquad\qquad
(\rho\to\infty)\,\,.
\eeq
\begin{figure}[ht]
\begin{center}
\includegraphics[width=0.95\textwidth]{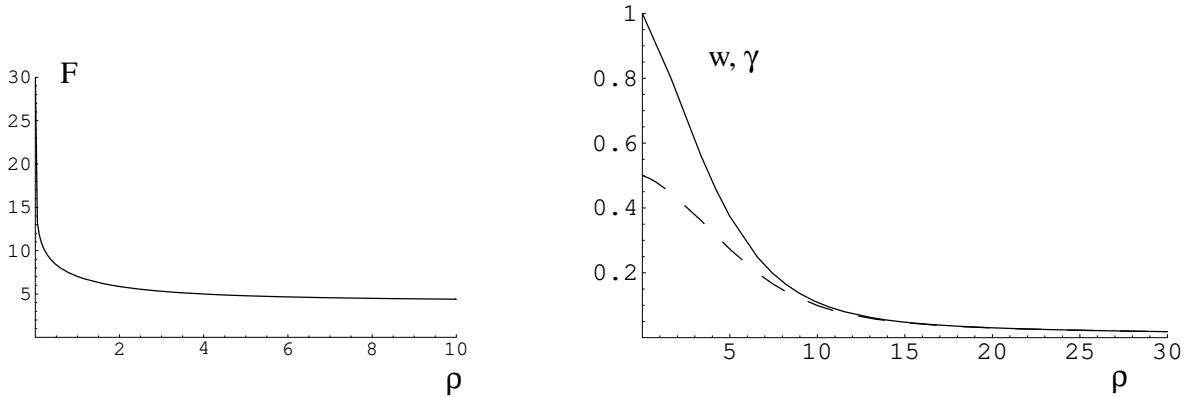}
\end{center}
\caption[Fwg]{On the left we plot $F$ as a function of the holographic coordinate $\rho$ for $N_c=4$, $N_f=1$. On the right the full (dashed) line corresponds to the function $w$($\gamma$) for the same values of $N_c$ and $N_f$.}
\label{Fwg}
\end{figure}

In order to find numerically the solution for $F$, $w$ and $\gamma$ for the full range of the holographic coordinate one has to match the IR regularity conditions (\ref{F-IR}), (\ref{initialw}) and (\ref{gamma-initial-flavored}) with the UV behavior (\ref{F-inseries-untruncated})-(\ref{bc-coefficients-untruncated}). As mentioned above, the only free parameter is $F_0=F(\rho=0)$. We have checked that such an interpolation between the
$\rho\to 0$ and $\rho\to \infty$ behaviors is possible by solving the BPS system with the IR initial conditions and by applying a shooting technique in which $F_0$ is varied until we obtain a solution with $F(\rho)\approx N_c$ for large $\rho$. This only happens when $F_0$ is fine-tuned to a very precise value. The result of this interpolation for $N_c=4$, $N_f=1$ is shown in figure \ref{Fwg}. The plot of $F(\rho)$ in this figure should be compared with the one in figure \ref{FPhiNc4Nf1}, which corresponds to the truncated system. The main difference between the results for $F(\rho)$ in these two figures is that $F(\rho)$ diverges for $\rho\to 0$ in the truncated system, while it remains finite in the complete solution, whereas for large $\rho$ both solutions nearly coincide. Notice that $w$ and $\gamma$ evolve smoothly from their initial values at $\rho=0$ to their vanishing asymptotic values for large $\rho$. The dilaton, which is not shown in figure \ref{Fwg}, grows monotonically with $\rho$ and becomes approximately a linear function of the holographic coordinate when $\rho$ is not very small. These features are in agreement with the expectation that these solutions of the complete system would reduce to the equivalent ones of the truncated ansatz  in the UV. Notice that in the borderline case $N_c=2N_f$ the initial value of $\gamma$ is zero. The result of our numerical calculation shows that, in this case, the function $\gamma$ becomes negative for $\rho>0$  and approaches its asymptotic vanishing value for $\rho\to\infty$ from negative values, in agreement with expansion written in (\ref{wgamma-inseries-untruncated}).

Having obtained this solution of the equations of motion of the gravity plus brane system, let us see if it incorporates some of the features that the supergravity dual of 2+1 dimensional gauge theory plus flavors should exhibit.  First of all, in the next subsection we will give a prescription to evaluate the gauge coupling and we will verify that, for $N_c\ge 2N_f$, this coupling displays the expected property of asymptotic freedom in the UV. Moreover, in subsection \ref{Wilsonloop-section} we will analyze the potential energy for an external quark-antiquark pair and we will discover that this potential behaves  in the way expected in a theory which has  string breaking due to pair production of dynamical massless quarks.

\subsubsection{The Yang-Mills coupling and the beta function}
Let us study the evolution of the gauge  coupling constant with the holographic coordinate. In order to do that, let us consider a D5-brane  probe extended along the three Minkowski directions and wrapping some internal three-cycle at a fixed value of the holographic coordinate. The natural three-cycle to compute the Yang-Mills coupling is just the one used above to fix the constant $\kappa$, namely $\Sigma=\{\omega^i=\sigma^i\}$. Indeed, as shown in section \ref{Unflavor-section}, $\Sigma$ shrinks to zero size at $\rho\to 0$, which corresponds to the IR of the gauge theory where one expects to have a 2+1 dimensional behavior of the D5-brane probe. Thus, $\Sigma$ is the analogue in the present case of the two-cycle found in ref. \cite{Bertolini:2002yr} for the background dual to ${\cal N}=1$ super Yang-Mills in four dimensions.

 \begin{figure}[ht]
\begin{center}
\includegraphics[width=0.4\textwidth]{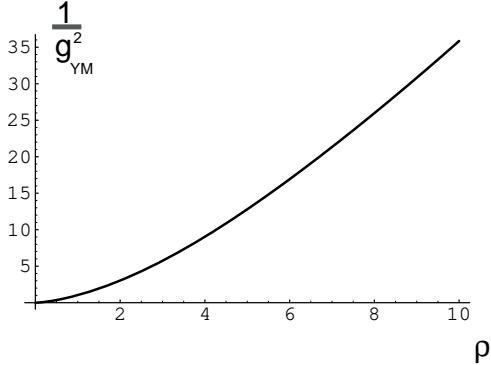}
\end{center}
\caption[gym]{The inverse of the Yang-Mills coupling as a function of the holographic coordinate $\rho$ for the case $N_c=4$, $N_f=1$.}
\label{gym}
\end{figure}
The DBI action for such a probe in the Einstein frame is:
\beq
S_{DBI}\,=\,-T_{D5}\,\,\int_{{\cal M}_6}\,\,
e^{{\phi\over 2}}\,\,\sqrt{-\det \Big(\,\hat G_6+e^{-{\phi\over 2}}{\cal F}\,\Big)}\,\,,
\eeq
where $\hat G_6$ is the induced metric on the D5-brane worldvolume and 
 ${\cal F}$ is the worldvolume gauge field. By looking at the ${\cal F}^2$ terms in the above action, we get the value of the Yang-Mills coupling constant of the dual 2+1 gauge theory, namely:
\beq
{1\over g^2_{YM}}\,=\,e^{-{3\over 4}\,\phi}\,\,
\int_{\Sigma}\,\,\sqrt{- \det \big(\,\hat G_3\,\big)}\,\, \,d^3\,\xi\,\,,
\eeq
where the induced metric $\hat G_3$ on the three-cycle $\Sigma$ has been written in eq. (\ref{ds-on-Sigma}) and we have  neglected   all constant  numerical factors. By using the metric written in  (\ref{ds-on-Sigma}), we obtain:
\beq
{1\over g^2_{YM}}\,=\,\Big[\,\rho+\,{F\over 4}\,(1-w)^2\,\Big]^{{3\over 2}}\,\,,
\label{gYM}
\eeq
where again we have neglected all numerical multiplicative constants.  Due to our boundary condition (\ref{initialw}), the right-hand side of (\ref{gYM}) vanishes for $\rho=0$, which corresponds to having $g^2_{YM}\to\infty$ in the IR, as expected in a confining theory. Clearly, $1/g^2_{YM}$ grows as we move towards the UV region $\rho\to\infty$, in agreement with the expected property of asymptotic freedom. In figure \ref{gym} we have plotted $1/g^2_{YM}$ for $N_c=4$ and $N_f=1$. In order to obtain the corresponding beta function from (\ref{gYM}) we would need the relation between the coordinate $\rho$ and the energy scale of the problem. The usual arguments employed for the gravity duals  of four-dimensional ${\cal N}=1$ gauge theory are not valid in our three-dimensional case and, as a consequence, such an energy-radius relation is lacking here. The best that we can do is to use the original radial variable $r$ which, as argued in subsection \ref{FMnas-abe-subsection}, grows in all cases when we move towards the UV. This fact can be further verified by placing a fundamental string stretched along the radial direction and looking at its energy as given by the Nambu-Goto action. Clearly this energy grows in the direction of increasing dilaton, \ie\ when $r$ increases. When $N_c\ge 2 N_f$, large $\rho$ corresponds to large $r$ and, as $F\to N_c$ and $w\to 0$ for our solutions, one approximately has:
\beq
{d\over dr}\,\Big(\,{1\over g^2_{YM}}\,\Big)\,\approx\,{3\over 2}\,\,
\rho^{{1\over 2}}\,\,{d\rho\over dr}\,\,,
\qquad\qquad
(\rho\to\infty)\,\,.
\label{beta-general}
\eeq
Moreover, when $N_c >2 N_f$ we get from the asymptotic expansion (\ref{r-rho-aympt}):
\beq
{d\rho\over dr}\,\sim {2(N_c-2N_f)\over \sqrt{N_c}}
\qquad\qquad
(\rho\to\infty)\,\,,
\eeq
and, thus, we can write for large $r$ and $\rho$:
\beq
{d\over dr}\,\Big(\,{1\over g^2_{YM}}\,\Big)\,\sim\,{(N_c-2N_f)
\over \sqrt{ N_c}}\,\,
\rho^{{1\over 2}}\,\sim\,
{(N_c-2N_f)^{{3\over 2}}\over (N_c)^{{3\over 4}}}\,\,r^{{1\over 2}}
\qquad\qquad
(N_c>2N_f)
\label{beta-up}
\eeq
Eq. (\ref{beta-up}) shows that $1/g^2_{YM}$ grows with $r$ in the UV (as it is obvious from figure \ref{gym}) and that its derivative also grows with $r$ as $\sqrt{r}$. This behavior is consistent with the expected negative beta function for $g^2_{YM}$. The form of eq. (\ref{beta-up}) could lead to the conclusion  that its right-hand side vanishes in the borderline case $N_c=2N_f$.  However, one should be careful  in this  case and use the correct expression (\ref{r-rho-border}) in (\ref{beta-general}). One gets for large $\rho$:
\beq
{d\over dr}\,\Big(\,{1\over g^2_{YM}}\,\Big)\,\sim\, {1\over \rho^{{1\over 2}}}\,\sim\,
{1\over r^{{1\over 4}}}\,\,,\qquad\qquad
(N_c=2N_f)\,\,,
\eeq
which shows that, actually, that the right-hand side only vanishes when $r\to\infty$ and the theory is still asymptotically free.

\subsubsection{Wilson loops}
\label{Wilsonloop-section}

In order to verify how the flavor degrees of freedom are encoded in our backreacted geometry, let us study the rectangular Wilson loops for external, non-dynamical,  heavy quarks.  These Wilson loops can be evaluated  by studying the Nambu-Goto action of a fundamental string whose ends lie in the UV region $r\to\infty$ and are separated by a distance $L$ in the gauge theory directions \cite{Maldacena:1998im,Rey:1998ik}(see also \cite{Brandhuber:1998er,Sonnenschein:1999if}). To describe such configurations let us choose the time $t$ and a  Minkowski coordinate $x$ as worldvolume coordinates and let us parameterize the string worldsheet by means of a function $r=r(x)$, where $r$ is the holographic coordinate of (\ref{ansatz}). The induced metric in the string frame is:
\beq
e^{\phi}\,[\,-dt^2\,+\,(\,1\,+\,(r\,')^2\,)\,dx^2\,]\,\,,
\eeq
and, as a consequence, the Nambu-Goto action takes the form:
\beq
S\,=\,-\int dt dx\,\, e^{\phi (r)}\,\sqrt{1\,+\,(r\,')^2}\,\,,
\label{NGaction}
\eeq
where we are taking the string tension to be equal to one and $r\,'=dr/dx$.  From the invariance of the lagrangian in (\ref{NGaction}) under shifts in the coordinate $x$, we immediately obtain a first integral of the equations of motion of the string, namely:
\beq
{e^{\phi (r)}\over \sqrt{1\,+\,(r\,')^2}}\,=\,e^{\phi (r_{min})}\,\,,
\label{NG-firstintegral}
\eeq
where $r_{min}$ is the minimal value of the holographic coordinate reached by the string worldsheet.  From (\ref{NG-firstintegral}) we can straightforwardly obtain $ r\,'$, with the result:
\beq
r\,'\,=\,\pm\,{\sqrt{
e^{2\phi (r)}\,-\,e^{2\phi (r_{min})}}\over e^{\phi (r_{min})}}\,\,.
\eeq
\begin{figure}[ht]
\begin{center}
\includegraphics[width=0.95\textwidth]{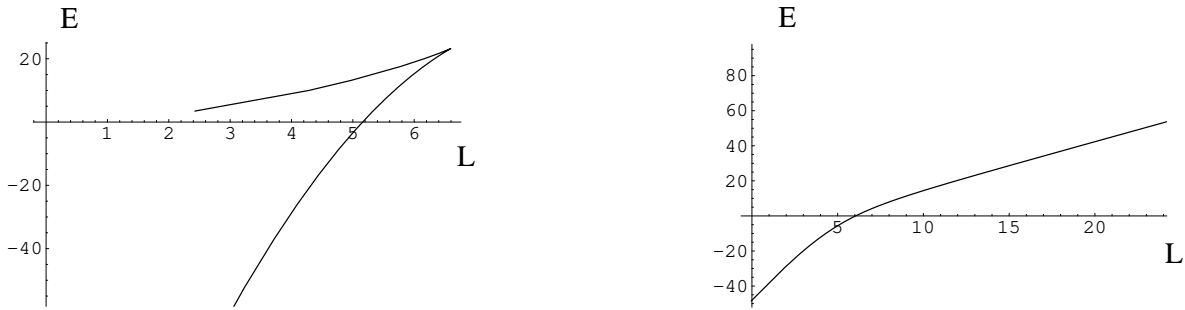}
\end{center}
\caption[Wilson]{In this figure we plot the energy $E$ of the Wilson loop versus the length  $L$. The curve on the left corresponds to the case $N_c=4$, $N_f=1$. This curve clearly shows that there is a maximal   $L$ and that $E$ is a double-valued function of $L$. On the right we plot the same quantity for the unflavored case with $N_c=4$, where string breaking does not take place.}
\label{Wilson}
\end{figure}

It is now trivial to obtain the length $L$, \ie\ the separation between the quark and the antiquark, as a function of the minimal value of $r$:
\beq
L(r_{min})\,=\,2\,\int_{r_{min}}^{r_{max}}\,\,
{e^{\phi (r_{min})}\over 
\sqrt{
e^{2\phi (r)}\,-\,e^{2\phi (r_{min})}}}\,\,dr\,\,,
\label{LWilson}
\eeq
where $r_{max}$ is a cutoff related to the mass of the external quarks, that can be taken to be very large. Moreover, after subtracting the masses of the non-dynamical quarks, the energy of the string configuration becomes:
\beq
E(r_{min})\,=\,2\,
\int_{r_{min}}^{r_{max}}\,\,
{e^{2\phi (r)}\over \sqrt{e^{2\phi (r)}\,-\,e^{2\phi (r_{min})}}}\,\,dr
\,-\,\int_{0}^{r_{max}}\,\,e^{2\phi (r)}\,\,dr\,\,,
\eeq
which can be identified with the potential energy of a quark-antiquark pair separated by a distance $L(r_{min})$.  Notice that both $E$ and $L$ depend parametrically on $r_{min}$. By varying $r_{min}$ we can obtain the corresponding  values of $E$ and $L$ and extract the  dependence of the energy on the distance. In a theory with dynamical quarks one expects the strings to elongate until their tension equals the mass of the lightest meson and then  create a quark-antiquark pair and break. In an $E$ versus $L$ plot this behavior would correspond to having a maximal value of $L$ and a double-valued function $E(L)$. The corresponding result for our solution  when $N_c=4$  and $N_f=1$  are displayed in figure \ref{Wilson}. We see that the  expected behavior of $E(L)$ is reproduced, in a way similar to one found in ref. \cite{Casero:2006pt} for the background dual to  an ${\cal N}=1$ SQCD-like theory in 3+1 dimensions. To allow for a comparison with the unflavored theory we have also plotted in figure \ref{Wilson} the result of the $E(L)$ curve for $N_c=4$  and $N_f=0$. In this unflavored case there is no maximal value of $L$ and, for large quark-antiquark separation,  the energy grows linearly with $L$, as it should for a confining theory without screening due to pair creation. 

It is also possible to understand the different behaviors of the $E(L)$ curves by analyzing the function that is integrated on the right-hand side of (\ref{LWilson}). Indeed, one can check that when $r$ and $r_{min}$ are both small, the square root on the denominator in (\ref{LWilson}) behaves as:
\beq
\sqrt{e^{2\phi (r)}\,-\,e^{2\phi (r_{min})}}\,\sim \, r^{\alpha}\,\,,
\qquad\qquad (r, r_{min})\to 0\,\,,
\eeq
where $\alpha$ is some constant. In the unflavored case, by combining eq. 
(\ref{IR-dilaton-untruncated-noflavor}) and the fact that $\rho\sim r^2$ for small $r$, one concludes that $\alpha=1$, which in turn implies that the integral giving $L(r_{min})$ is divergent when $r_{min}\to 0$, in agreement with our numerical results. On the contrary, when flavors are added one can verify numerically that $\alpha<1$, which means that the integral (\ref{LWilson}) is  now convergent when $r_{min}\to 0$ and there is a maximal value of $L$. This different behavior of the dilaton in these two cases is correlated with the fact that the flavored metric develops a 
(good) curvature singularity at the origin, while the unflavored solution is regular. It is worth pointing out that similar results have been found in \cite{Casero:2006pt}.

\subsection{Asymptotic $G_2$ cones}
\label{G2cones-untruncated-subsection}

When $F_0$ takes values in a certain range, the solutions of the BPS equations reached at the UV have a metric which is the direct product of 2+1 dimensional Minkowski space and a $G_2$ cone. The solutions in this case are very similar to the ones discussed in subsections \ref{explicitsolution-unflavored-section} and \ref{G2cones-truncated-subsection} and we will not discuss them further here. Let us only mention that the asymptotic values of $F$, $w$ and $\gamma$  for $\rho\to\infty$ can be determined analytically, in a way completely analogous  to the one employed in subsection \ref{explicitsolution-unflavored-section} (one has to take into account the different value (\ref{kappa-flavors}) of  the constant $\kappa$ in the present flavored case). One gets the following asymptotic behavior:
\bear
&&F\,\approx\,{4\over 3}\,\rho\,+\,4\,(\,N_f-N_c)\,+\,\cdots\,\,,\rc\rc
&&w\,\approx\,{3(N_c-3N_f)\over 2\rho}\,+\,\cdots\,\,,
\qquad\qquad (\rho\to\infty)\,\,,
\rc\rc
&&\gamma\,\approx\,{1\over 3}\,-\,{N_f\over N_c}\,+\,\cdots\,\,\,,
\eear
a result that is confirmed by our numerical calculation.

\section{The general system with flavor ($N_c<2N_f$)}
\label{untruncated-flavor-Landau}

In this section we briefly describe the solutions of the complete BPS system for $N_c<2N_f$ that have an asymptotic linear dilaton in the UV. Even if the interpretation of these solutions is less clear than in  the $N_c\ge 2N_f$ case, it follows from the analysis performed in subsection \ref{FMnas-abe-subsection} for the truncated solutions that the right holographic variable is now the original coordinate $r$, instead of the variable $\rho$ used so far. Moreover, we learned in that subsection that, in this $N_c<2N_f$ case, the small $\rho$ region should be interpreted as the UV of the gauge theory (with large $r$) and vice versa, the IR would correspond to large $\rho$ and small $r$. Somehow when going from $N_c\ge 2N_f$ to $N_c< 2N_f$ the UV and IR are exchanged.  Thus, when $N_c< 2N_f$, it is natural to search for solutions that approach  the unflavored one at the IR\footnote{This is actually what happens in other backreacted solutions with a Landau pole, such as the ones in \cite{Benini:2006hh} for the conifold.}
 where, in terms of the variable $\rho$, they can be represented by the series (\ref{F-inseries-untruncated}) and (\ref{wgamma-inseries-untruncated}). Notice that when one uses the variable $r$ as the independent variable one should also determine the function
$\rho(r)$ or, equivalently, $h=h(r)$, where $h(r)$ is the function squashing the $\sigma^i$ sphere in our ansatz (\ref{ansatz}). The differential equation determining $\rho(r)$ is just the one written in (\ref{drdho-untruncated}).

\begin{figure}[ht]
\begin{center}
\includegraphics[width=0.40\textwidth]{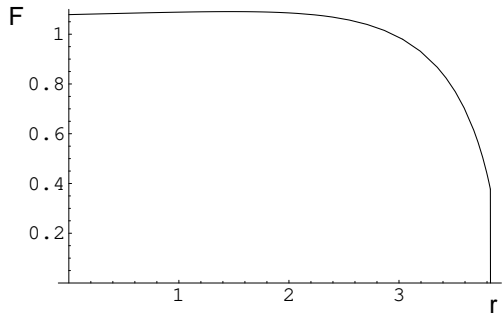}
\end{center}
\caption[NcNfone]{The function $F(\rho)$ for $N_c=N_f=1$. To obtain this curve we have imposed the behavior (\ref{F-inseries-untruncated}) for small $r$ (or large $\rho$).
}
\label{NcNfone}
\end{figure}

The numerical  integration of the BPS system for $N_c<2N_f$ with the IR initial conditions  given by (\ref{F-inseries-untruncated}) and (\ref{wgamma-inseries-untruncated}) shows that the function $F$ decreases from its initial value $F\approx N_c$ at $r\approx 0$ until it vanishes at some finite value $r_*$ of the coordinate  $r$. For larger values of $r$ the function $F$ becomes negative and the solution does not make sense any more. In figure \ref{NcNfone} we plot $F$ as a function of $r$ for $N_c=N_f=1$. Notice that $F(r)$ drops very fast to zero as $r$ approaches its final value $r_*$. By choosing appropriately the initial conditions in the integration of $\rho(r)$ (or, equivalently,  of $h(r)$) we can make that $r=r_*$ corresponds to $\rho=0$. 
Moreover, the dilaton (not shown in figure \ref{NcNfone}) grows linearly with $r$, as expected for this type of solutions.  In these calculations we have taken the same value (\ref{kappa-flavors}) for the constant $\kappa$. Notice that, with our choice of initial conditions, the cycle $\Sigma$ collapses at $r=r_*$.

As argued in subsection \ref{FMnas-abe-subsection} for the truncated system, we think that the proper interpretation of the point $r=r_*$  is the location of a Landau pole. To confirm this interpretation we have calculated the Yang-Mills coupling, as given by (\ref{gYM}), as a function of the radial variable $r$. The numerical results  confirm that $1/g^2_{YM}$ is a monotonically decreasing function of $r$ and that  $g^2_{YM}\to \infty$ as we approach the point $r=r_*$.

\section{Summary and discussion}
\label{conclusions}

In this paper we have found  backgrounds which encode the effect of adding a large number of unquenched flavors to the gravity dual of ${\cal N}=1$ gauge theories in 
 2+1 dimensions. By using kappa symmetry, we first determined the appropriate embeddings of the flavor branes that preserve the supersymmetries of the unflavored background and, then we found  the modification of the ansatz of the RR field needed to solve the Bianchi identity of the coupled gravity plus branes system. We have subsequently obtained a system of first-order BPS equations, which we have solved with different boundary conditions. The most interesting solutions are those that contain a linear dilaton in the UV. When $N_c\ge 2N_f$ we have argued that these solutions display the expected properties of a gravity dual of an asymptotically free theory with dynamical quarks.  We have checked this fact by computing, from our background, the Yang-Mills coupling constant, as well as the expectation value of the Wilson loop. In this latter case we have explicitly verified the expected string breaking due to  quark-antiquark pair production.  We also found solutions for $N_c<2N_f$ and we have shown that they are consistent with having a Landau pole in the UV of the gauge theory.
 
Let us comment on some points that, in our opinion, would need some further clarification. First of all, it would be desirable to have a more precise characterization of the field theory dual to the  background found here. One could argue as in \cite{Casero:2006pt} and try to determine the IR field theory that is obtained by integrating out the massive KK fields. Following the reasoning of \cite{Casero:2006pt} we conclude that extra couplings of the fundamental matter fields  (quartic or with higher powers) are generated. For this reason some of our  results are difficult to check on the field theory side. 

One of the problems that   would be interesting to understand from the field theory point of view is the dependence of the beta function on $N_c$ and $N_f$. Notice that, due to the low amount of supersymmetry preserved by our solution, one cannot rely on the power of holomorphy, which has been so useful to extract the non-perturbative structure of ${\cal N}=1$ gauge theories in four dimensions. In a certain sense these theories are a good arena to test the power of holography as a tool to explore the strong coupling regime of gauge theories. As a first step in this direction, let us try to determine the coefficient $k$ of the Chern-Simons term for our solution. In general, to obtain such a result one should be able to find the shift of the level due to the integration out of the KK fermions. In the presence of massless flavors this calculation is even more complicated  because an explicit computation of the Witten index is, at least to our knowledge, lacking. However, the form of the supergravity solution that we find seems to suggest that the Chern-Simons level $k$ of the low-energy three dimensional theory, after having integrated out the KK fermions, is:
\beq
k={N_c-3 N_f\over 2}\,\,.
\eeq
This would imply the Witten index for such theory would be:
\begin{equation}
I(n)\,=\,\frac{(N_c+n)!}{n!N_c!}
\end{equation}
with $n=k-(N_c-3N_f)/2\geq 0$, and being $0$ otherwise. It would be very interesting to 
prove or disprove these results from a direct calculation in  the field theory. 

Let us finish this section by mentioning some further topics that one could address from the supergravity side. First of all, one could try to generalize our background to the case in which the flavors are massive. For the case of the conifold  this generalization was achieved in \cite{Benini:2006hh} by a simple modification of the RR form whose Bianchi identity is violated. One could try to apply a similar procedure for the setup studied here. The analysis of the meson spectra for the backreacted geometry is clearly another interesting problem to look at. To perform this analysis one can add a probe and consider its fluctuations. Presumably one would find the same type of problems related to the normalizability of the fluctuation modes as in other backgrounds generated  by D5-branes and a careful treatment would be needed to extract the meson masses.  The construction of the black hole version of our background is clearly of great interest since it would allow us to explore the thermodynamic and hydrodynamic properties of the field theory dual at finite temperature.  

For the metrics with $G_2$ holonomy  one can perform  the so-called flop transformation, in which the two three-spheres are exchanged. An interesting problem to study is to what extent this transformation can also be performed in our solutions and what are the effects on the field theory side (see \cite{Edelstein:2002zy} for a similar study in the case of backgrounds of $G_2$ holonomy without fluxes). According to the ideas put forward in \cite{Casero:2006pt}, one expects  that this discrete transformation  is a kind of Seiberg duality, that somehow would exchange the rank and the level of the corresponding field theory. 
Finally, one could try to see if our family of unflavored backgrounds can be used to describe the physics of domain walls in ${\cal N}=1$ gauge theories in four dimensions.

\section*{Acknowledgments}

We are grateful to D. Are\'an, S. Cremonesi, C. N\'u\~nez, A. Paredes, D. Rodriguez-G\'omez,  and  J. Shock for discussions and encouragement. 
This  work was supported in part by MEC and  FEDER  under grant
FPA2005-00188,  by the Spanish Consolider-Ingenio 2010 Programme CPAN (CSD2007-00042), by Xunta de Galicia (Conselleria de Educacion and grant PGIDIT06PXIB206185PR)
and by  the EC Commission under  grant MRTN-CT-2004-005104.

\vskip 1cm
\renewcommand{\theequation}{\rm{A}.\arabic{equation}}
\setcounter{equation}{0}
\appendix
\section{Appendix: Derivation of the BPS equations}
\label{BPSapp}

The supersymmetry transformations for the type IIB dilatino $\lambda$  and 
gravitino $\Psi_{\mu}$  in Einstein frame, when the RR three-form is nonzero, are:
\bear
\delta\lambda&=&{i\over
2}\,\,\partial_{\mu}\,\phi\,\Gamma^{\mu}\,\epsilon^*\,+\, {1\over
24}\,e^{{\phi\over 2}}\,
F^{(3)}_{\mu_1\mu_2\mu_3}\,\Gamma^{\mu_1\mu_2\mu_3}\,\epsilon\,\,,\rc\rc
\delta\Psi_{\mu}&=&D_{\mu}\epsilon\,+\,{i\over 96}\,
e^{{\phi\over 2}}\,F^{(3)}_{\mu_1\mu_2\mu_3}\,
\Big(\,\Gamma_{\mu}^{\,\,\mu_1\mu_2\mu_3}\,-9\,\delta_{\mu}^{\mu_1}\,
\Gamma^{\mu_2\mu_3}\,\Big)\,\epsilon^*\,\,.
\label{SUSYIIB}
\eear
We want to solve the conditions $\delta\lambda=\delta\psi_{\mu}=0$ for a metric given by our ansatz (\ref{ansatz}). In what follows we shall 
 choose the following vierbein basis:
\bear
&&e^{x^i}\,=\,e^f\,dx^i\,\,,\qquad \qquad e^r\,=\,e^f\,dr\,\,,\rc\rc
&&e^i\,=\,e^{f+h}\,{\sigma^i\over 2}\,\,,\qquad \qquad 
e^{\hat i}\,=\,e^{f+g}\,\Big(\,{\omega^i-A^i\over 2}\,\Big)\,\,,
\label{basis}
\eear
where $A^i$ is parameterized in terms of the function $w(r)$ as in (\ref{Ai-w}). 
The spin connection in the basis (\ref{basis})  is given by:
 \bear
&&\omega^{x^i r}\,=\,e^{-f}\,f'\,e^{x^i}\,\,,\rc
&&\omega^{ir}\,=\,e^{-f}\,(f'\,+\,h')\,e^{i}\,-\,{w'\over 4}\,e^{-f+g-h}\,e^{\hat i}\,\,,\rc
&&\omega^{\hat i r}\,=\, e^{-f}\,(f'\,+\,g')\,e^{\hat i}\,-\,
{w'\over 4}\,e^{-f+g-h}\,e^{ i}\,\,,\rc
&&\omega^{ij}\,=\,-\epsilon^{ijk}\,\Big[\,e^{-f-h}\,e^{k}\,+\,{1-w^2\over 4}\,\,
e^{-f+g-2h}\,e^{ \hat k}\,\Big]\,\,,\rc
&&\omega^{\hat i \hat j}\,=\,-\epsilon^{ijk}\,\Big[\,e^{-f-g}\,e^{\hat k}\,+\,(1+w)\,
e^{-f-h}\,e^{k}\,\Big]\,\,,\rc
&&\omega^{ i \hat j}\,=\,{1-w^2\over 4}\,\,\epsilon^{ijk}\,
e^{-f+g-2h}\,e^{  k}\,-\,{w'\over 4}\,e^{-f+g-h}\,\delta_{ij}\,e^{r}\,\,,\rc
&&\omega^{ \hat i j}\,=\,{1-w^2\over 4}\,\,\epsilon^{ijk}\,
e^{-f+g-2h}\,e^{  k}\,+\,{w'\over 4}\,e^{-f+g-h}\,\delta_{ij}\,e^{r}\,\,.
\eear

We will take the RR three-form $F_3$ that corresponds to the general system with flavor of the main text, \ie\ we will take $F_3$ as given by (\ref{F3f3}).
The different components of this field strength in the basis (\ref{basis}) are:
\begin{eqnarray}
F_{ri\hat{i}}^{(3)}&=&\frac{N_c}{2}\gamma'e^{-3f-g-h}\,\,,\rc
 F_{\hat{1}\hat{2}\hat{3}}^{(3)}&=&-2N_ce^{-3f-3g}\,\,,\rc
 F_{\hat{i}jk}^{(3)}&=&-{\epsilon_{ijk}\over 2}\,
N_c\,\Big(1\,-\,{4N_f\over N_c}\,
+\,w^2\,-\,2w\gamma\,\Big)\,e^{-3f-g-2h}\,\,,\rc
 F_{\hat{i}\hat{j}k}^{(3)}&=&-\epsilon_{ijk}N_c(w-\gamma)e^{-3f-2g-h}\,\,,\rc
 F_{123}^{(3)}&=&\frac{N_c\,V}{4}\,e^{-3f-3h}\,\,,
\end{eqnarray}
where $V$ is a function of $r$ defined as:

\beq
V\,=\,(w-3\gamma)\,(1-w^2)\,-\,4\Big(\,1\,-{3N_f\over N_c}\,\Big)\,
 w\,+\,8\kappa\,\,,
 \label{V-general}
\eeq
with $\kappa$ being the same  constant as in (\ref{calH}).

The Killing spinors of the background are those $\epsilon$ for which the right-hand side
of eq.  (\ref{SUSYIIB}) vanishes.  In order to satisfy the equations 
$\delta\lambda=\delta\Psi_{\mu}=0$ we will have to impose certain projection conditions
on $\epsilon$.  These conditions are:
\begin{eqnarray}
\Gamma_{1\hat{1}}\,\epsilon=\Gamma_{2\hat{2}}\,\epsilon=\Gamma_{3\hat{3}}\epsilon,\ \ \ \qquad i\epsilon^{\star}=\epsilon\,\,.
\label{projections}
\end{eqnarray}
Let us now define the following matrix
\beq
\Gamma_*\,\equiv\,\Gamma_{r\hat{1}\hat{2}\hat{3}}\,\,.
\eeq
From the vanishing of the dilatino variation under SUSY, we get:   
\bear
&&\phi'\epsilon\,-\,N_c\,\Big[\,e^{-3g}\,-\,{3\over 4}\,
\Big(\,1\,-\,{4N_f\over N_c}\,
+\,w^2\,-\,2w\gamma\,\Big)\,e^{-g-2h}\,
\Big]\,\Gamma_*\epsilon\,+\,{3\over 4}\,N_c\,e^{-g-h}\,\gamma'\,
\,\Gamma_{1\hat 1}\,\epsilon\,-\rc\rc
&&\qquad\qquad
-{N_c\over 2}\Big[\,{V\over 4}\,e^{-3h}\,+\,3(w-\gamma)\,e^{{\phi\over 2}\,-\,2f-2g-h}
\,\Big]\,\Gamma_{1\hat 1}\,\Gamma_*\epsilon\,=\,0\,\,.
\label{dilatino}
\eear
Let us now consider the gravitino variation. From the condition $\delta\psi_{x^i}=0$ we obtain that the metric function $f$ must be related to the dilaton as:
\beq
f\,=\,{\phi\over 4}\,\,.
\label{f-phi}
\eeq
Moreover, from the equation $\delta\psi_{i}=0$, after using eqs. (\ref{dilatino}) and (\ref{f-phi}), one arrives at:
\bear
&&h'\epsilon\,-\,{e^{-2h}\over 2}\,\Big[\,(1-w^2)\,e^{g}\,+\,N_c\,
\Big(\,1\,-\,{4N_f\over N_c}\, +\,w^2\,-\,2w\gamma\,\Big)\,e^{-g}\,
\Big]\Gamma_*\epsilon\,-\,\rc\rc
&&\qquad-\,{e^{g-h}\over 4}\,\Big[\,w'\,+\,N_c\,e^{-2g}\,\gamma'\,\Big]\,
\Gamma_{1\hat{1}}\,\epsilon\,+\,\rc\rc
&&\qquad\,+\,
{1\over 2}\,\Big[\,{N_c\over 4}Ve^{-3h}\,+\,N_c\,(w-\gamma)e^{-2g-h}
\,-\,2w\,e^{-h}\,\Big]\,
\Gamma_{1\hat{1}}\,\Gamma_*\,\epsilon\,=\,0\,\,.
\label{psi-i}
\eear
Similarly, the condition $\delta\psi_{\hat i}=0$ leads to:
\bear
&&g'\epsilon\,+\,{1\over 4}\,\Big[\,(1-w^2)\,e^{-2h+g}-4e^{-g}-N_c\,\Big(
\Big(\,1\,-\,{4N_f\over N_c}\, +\,w^2\,-\,2w\gamma\,\Big)\,e^{-g-2h}\,-\,4e^{-3g}
\,\Big)\Big]\,\Gamma_*\epsilon\,+\,\rc\rc
&&\qquad\qquad
+\,{e^{g-h}\over 4}\,\Big[\,w'\,-\,N_c\,e^{-2g}\,\gamma'\,\Big]\,
\Gamma_{1\hat{1}}\,\epsilon\,+\,N_c\,(w-\gamma)\,e^{-2g-h}\,\Gamma_{1\hat 1}\,
\Gamma_*\,\epsilon\,=\,0\,\,.
\label{psi-hat-i}
\eear
In order to solve the above equations, we shall impose the additional projection:
\begin{equation}
\Gamma_{*}\epsilon=(\beta+\tilde{\beta}\Gamma_{1\hat{1}})\epsilon\,\,,
\label{Gamma*}
\end{equation}
where $\beta$ and $\tilde\beta$ are functions of the radial variable to be determined. As
$(\Gamma_*)^2=1$ and $\{\Gamma_{*}, \Gamma_{1\hat{1}}\}=0$, by consistency, these quantities must satisfy the condition:
\beq
\beta^2\,+\,\tilde\beta^2\,=\,1\,\,,
\label{beta-normalization}
\eeq
and, therefore, they can be represented in terms of a single angle $\alpha$ as:
\begin{equation}
\beta=\cos\alpha\,\,,\qquad\qquad
 \tilde{\beta}=\sin\alpha\,\,.
 \label{betas-alphas}
\end{equation} 
Notice that the projection (\ref{Gamma*}) is equivalent to
\begin{equation}
\Gamma_{1\hat{1}}\,\Gamma_{*}\,\epsilon=(\beta\Gamma_{1\hat{1}}-\tilde{\beta})\,
\epsilon\,\,,
\label{Gamma1Gamma*}
\end{equation}
and can be solved as:
\begin{equation}
\epsilon=e^{-\frac{\alpha}{2}\Gamma_{1\hat{1}}}\,\epsilon_0\label{alpha}\,\,,
\end{equation}
where $\alpha$ and the spinor $\epsilon_0$ depend on $r$ and the latter satisfies  the projection:
\begin{equation}
\Gamma_{*}\,\epsilon_0=\epsilon_0.
\end{equation}

We can now write the set of BPS equations corresponding to this ansatz. Let us substitute (\ref{Gamma*}) and (\ref{Gamma1Gamma*}) in the dilatino variation
(\ref{dilatino}). By separating the terms containing the unit matrix from those with
 $\Gamma_{1\hat 1}$,  and using $\phi=4f$, we arrive at the  following equations:
\begin{eqnarray}
f'&=&{N_c\over 4}\left[e^{-3g}-\frac{3}{4}\Big(\,1\,-\,{4N_f\over N_c}\, +\,w^2\,-\,2w\gamma\,\Big)\,e^{-g-2h}\,\right]\beta
-\frac{N_c}{8}\left[\frac{V}{4}e^{-3h}+3(w-\gamma)e^{-2g-h}\right]\tilde\beta\,\,,
\rc\rc
\gamma'&=&\frac{4}{3}\left[e^{-2g+h}-
\frac{3}{4}\Big(\,1\,-\,{4N_f\over N_c}\, +\,w^2\,-\,2w\gamma\,\Big)
e^{-h}\right]\tilde{\beta}
+\frac{2}{3}\left[\frac{V}{4}e^{-2h+g}+3(w-\gamma)e^{-g}\right]\beta\,\,.\rc\rc
\label{phi-gamma}
\end{eqnarray}
In order to determine $\beta$ and $\tilde\beta$, let us plug (\ref{Gamma*}) and 
(\ref{Gamma1Gamma*}) in (\ref{psi-i}) and consider the terms containing 
$\Gamma_{1\hat 1}$. One gets:
\bear
&&e^{g-h}\,w'\,+\,N_c\,e^{-g-h}\,\gamma'\,=\,
\Big[\,{N_c\over 2}\,V\,e^{-3h}\,+\,2N_c\,(w-\gamma)\,e^{-2g-h}
-\,4w e^{-h}\,\Big]\beta
\,+\,\rc\rc
&&\qquad
+\,2\Big[\,(w^2-1)\,e^{g-2h}\,-\,N_c\,
\Big(1-{4N_f\over Nc}\,+\,w^2-2w\gamma\Big)\,e^{-g-2h}\,\Big]\,
\tilde\beta\,\,.
\label{w+gamma}
\eear
Similarly, from (\ref{psi-hat-i}) one arrives at:
\bear
&&e^{g-h}\,w'\,-\,N_c\,e^{-g-h}\,\gamma'\,=\,-4N_c(w-\gamma)\,e^{-2g-h}\,\beta\,-\,\rc\rc
&&-\,
\Big[\,(1-w^2)\,e^{g-2h}\,-\,4e^{-g}\,+\,4N_c e^{-3g}\,-\,N_c\,
\Big(1\,-\,{4N_f\over N_c}+w^2-2w\gamma\Big)\,e^{-g-2h}\,\Big]\,\tilde\beta\,\,.\rc
\label{w-gamma}
\eear
By substituting the expression of $\gamma'$ taken from (\ref{phi-gamma}) into (\ref{w+gamma}), one gets the following expression of $w'$:
\bear
&&e^{g-h}w'=\Big[\,{N_c\over 3}\,V\,e^{-3h}\,-\,4we^{-h}\,\Big]\beta\,-\,\rc\rc
&&\qquad
-\Big[\,2(1-w^2)\,e^{g-2h}+{4\over 3}\,N_c e^{-3g}+
N_c \Big(1\,-\,{4N_f\over N_c}+w^2-2w\gamma\Big) e^{-g-2h}\,\Big]\tilde\beta\,\,,\rc
\label{w-prime-1}
\eear
whereas, by performing  a similar manipulation to (\ref{w-gamma}), one can prove that:
\bear
&&e^{g-h}w'\,=\,N_c\,\Big[\,{V\over 6}\,e^{-3h}\,-\,2(w-\gamma)\,e^{-2g-h}\,\Big]
\beta+\rc\rc
&&\qquad\qquad
+\,\Big[\,4e^{-g}\,-\,{8\over 3}\, N_c\,e^{-3g}\,-\,(1-w^2)\,e^{g-2h}\,\Big]\tilde\beta\,\,.
\label{w-prime-2}
\eear
By eliminating $w'$ in (\ref{w-prime-1}) and (\ref{w-prime-2}), one can demonstrate 
that $\beta$ and $\tilde\beta$ satisfy a relation of the form:
\beq
{\tilde \Lambda }\,\beta\,-\,\Lambda\tilde\beta\,=\,0\,\,,
\label{calAB}
\eeq
where the functions $\Lambda$ and ${\tilde \Lambda}$ are given by:
\bear
&&\Lambda\,=\,e^{2h}\,+
{1-w^2\over 4}\,e^{2g}\,+\,{N_c\over 4}\,
\Big(1\,-\,{4N_f\over N_c}\,+\,w^2-2w\gamma\Big)\,-\,{N_c\over 3}\,e^{2h-2g}\,\,,\rc\rc
&&{\tilde \Lambda}\,=\,{N_cV\over 24}\,e^{g-h}\,-\,we^{g+h}\,+\,{N_c\over 2}\,(w-\gamma)\,e^{-g+h}
\,\,.
\label{Lambdas}
\eear
One can solve (\ref{calAB}) for $\beta$ and $\tilde\beta$ as $\beta\,\propto\,\Lambda$, 
$\,\,\tilde\beta\,\propto\,\tilde\Lambda$, where the common proportionality function  is determined by imposing the condition
$\beta^2+\tilde{\beta}^2=\sin^2\alpha+\cos^2\alpha=1$ (see eq. (\ref{beta-normalization})). One gets:
\beq
\beta\,=\,{\Lambda\over \sqrt{\Lambda^2\,+\,{\tilde \Lambda}^2}}\,\,,
\qquad\qquad
\tilde\beta\,=\,{\tilde \Lambda\over \sqrt{
\Lambda^2\,+\,{\tilde \Lambda}^2}}\,\,.
\label{betas-Lambdas}
\eeq

Let us now  rewrite (\ref{w-prime-2}) as:
\bear
&&w'\,=\,N_c\,\left[\frac{V}{6}e^{-g-2h}-2(w-\gamma)e^{-3g}\right]\beta\,+\,\rc\rc
&&\qquad\qquad
+\,\Big[\,4e^{-2g+h}\,-\,{8\over 3}\, N_c\,e^{-4g+h}\,-\,(1-w^2)\,e^{-h}\,\Big]\tilde\beta\,\,.
\label{BPS-w}
\eear
Up to now we have determined the first-order BPS equations satisfied by $f$, $\gamma$ and $w$ (eqs. (\ref{phi-gamma}) and (\ref{BPS-w})). Let us now determine the equations for the remaining 
functions of our ansatz, namely $h$ and $g$. By using (\ref{Gamma*}) and (\ref{Gamma1Gamma*}) in (\ref{psi-i}) and considering the terms containing the unit matrix , we get:
\begin{eqnarray}
&&h'=\frac{1}{2}\left[(1-w^2)e^{g-2h}\,+\,
N_c\Big(1-\,\frac{4N_f}{N_c}+w^2-2w\gamma\Big )e^{-g-2h}\right]
\,\beta\,+\rc\rc
&&\qquad\qquad
+\,\frac{1}{2}\left[{N_c\over 4}\,V\,e^{-3h}\,+\,
N_c\,(w-\gamma)e^{-2g-h}-2we^{-h}\right]\,\tilde\beta\,\,.
\label{hprime}
\end{eqnarray}
Similarly,  from (\ref{psi-hat-i}) we obtain the following equation for $h$:
\bear
&&g'=\frac{1}{4}\left[\,4e^{-g}\,-\,(1-w^2)e^{g-2h}\,+\,
N_c\,\Big(1-4\,\frac{N_f}{N_c}+w^2-2w\gamma\,\Big)e^{-g-2h}\,-\,4N_c\,e^{-3g}\,
\right]\beta\,+\,\rc\rc
&&\qquad\qquad\qquad\qquad\qquad\qquad\qquad
+N_c\,(w-\gamma)e^{-2g-h}\,\tilde{\beta}.
\label{gprime}
\eear

To complete our analysis of the Killing spinor equations we should check that the variation of the radial component of the gravitino vanishes. Actually, one can check that this condition holds if the following two equations are satisfied:
\begin{eqnarray}
\alpha'&=&-\frac{3}{4}N_c\gamma'e^{-g-h}-\frac{3}{4}w'e^{g-h}\,\,,\label{alpha-prime}\\
\epsilon_0'&=&\frac{\phi'}{8}\epsilon_0\,\,,
\label{epsilon0-prime}
\end{eqnarray}
where $\alpha$ is related to $\beta$ and $\tilde\beta$ as in (\ref{betas-alphas}). By using  (\ref{betas-Lambdas}) one can check, nontrivially,  that eq. (\ref{alpha-prime}) is a consequence of the other
first-order equations of the system. Moreover, (\ref{epsilon0-prime}) determines the dependence of $\epsilon_0$ on the radial coordinate. Indeed, it can be integrated as:
\beq
\epsilon_0\,=\,e^{{\phi\over 8}}\,\eta\,\,,
\label{epsilon0-eta}
\eeq
where $\eta$ is a constant spinor.
By combining  eqs. (\ref{epsilon0-eta}) and (\ref{alpha}), one can get the explicit expression of  the Killing spinors, namely:
\beq
\epsilon\,=\,e^{-\frac{\alpha}{2}\Gamma_{1\hat{1}}}\,e^{{\phi\over 8}}\,\eta\,\,,
\eeq
with $\eta$ being a constant spinor satisfying the four commuting projections:
\bear
&&\Gamma_{1\hat{1}}\,\eta=\Gamma_{2\hat{2}}\,\eta=\Gamma_{3\hat{3}}\,\eta,\rc
&& i\eta^{\star}=\eta\,\,,\rc
&&\Gamma_{*}\,\eta=\eta\,\,.
\label{projections-eta}
\eear
The projections (\ref{projections-eta}) show that our background is $1/16$ supersymmetric, \ie\ it preserves two supersymmetries. Notice that this is the amount of supersymmetry expected for an ${\cal N}=1$ theory in 2+1 dimensions. 

To finish this appendix, let us point out that the BPS equations just found are consistent with the truncation $w=\gamma=\kappa=0$ (which implies that $V=0$). It is clear from 
(\ref{alpha-prime}) that in this case the phase $\alpha$ can be taken to vanish or, equivalently, $\beta=1$ and $\tilde\beta=0$. The corresponding truncated equations are much simpler than the full BPS system and will be studied separately in the main text.

\section{Appendix: Equations of motion}
\label{EOMapp}
\renewcommand{\theequation}{\thesection.\arabic{equation}}
\setcounter{equation}{0}

The equation of motion for the dilaton derived from the action (\ref{totalS})  is:
\beq  
\frac{1}{\sqrt{-G}} \partial_M \Big ( G^{MN}\, \sqrt{-G}\, 
\partial_N \, \phi  \Big )\,-\,\frac{1}{12} e^{\phi} F^2_{3}\,=\,-\,
\frac{2\kappa^2_{10}}{\sqrt{-G}} \frac{\delta}{\delta \phi} S_{flavor}\,\, .
\label{dilaton-eom}
\eeq
In order to make this equation more explicit, let us compute the determinant of the metric  for our ansatz. The easiest way to do that is by realizing that $\sqrt{-G}$ can be obtained by computing the wedge product of all the one-forms of the frame basis:
\beq
e^{x^0}\wedge\cdots \wedge e^{\hat 3}\,=\, \sqrt{-G}\,\, dx^0\wedge\cdots dx^9\,\,.
\eeq
In our case, we get:
\beq
\sqrt{-G}\,=\,{1\over 64}\,\,e^{{5\phi\over 2}\,+\,3(h+g)}\,\,
\sqrt{\tilde g_1}\,\,\sqrt{\tilde g_2}\,\,,
\eeq
where $\tilde g_1$ and $\tilde g_2$ are angular factors that depend on the parametrization of the two sets of left-invariant $SU(2)$ one-forms. Actually, let us suppose that we represent the $\sigma^i$'s ($\omega^i$'s) in terms of the three angles
$\theta_1$, $\phi_1$ and $\psi_1$ ($\theta_2$, $\phi_2$ and $\psi_2$). Then:
\beq
\sigma^1\wedge\sigma^2\wedge\sigma^3\,=\,\sqrt{\tilde g_1}\,\,
d\theta_1\wedge d\phi_1\wedge d\psi_1\,\,,
\qquad
\omega^1\wedge \omega^2\wedge \omega^3\,=\,\sqrt{\tilde g_2}\,\,
d\theta_2\wedge d\phi_2\wedge d\psi_2\,\,.
\eeq
For the standard election of the angular variables, one can verify that:
\beq
\sqrt{\tilde g_i}\,=\,\sin\theta_i\,\,,\qquad\qquad (i=1,2)\,\,.
\eeq
Using these results one can verify that the equation of motion for the dilaton (\ref{dilaton-eom}) reduces to:
\beq
\phi''\,+\,2\Big(\,\phi'\,+\,3(h'\,+\,g')\Big)\phi'\,-\,{e^{{3\phi\over 2}}\over 12}\,\,F_3^2\,=\,
-\,\frac{2\kappa^2_{10}}{\sqrt{-G}} \,\,e^{{\phi\over 2}}\,\,
\frac{\delta}{\delta \phi} S_{flavor}\,\, .
\label{dil-eom}
\eeq
We shall evaluate the first two terms on the left-hand side of the above equation by using the BPS equations for the functions of the ansatz and the dilaton. Notice that, in order to compute the second derivative of the dilaton, we must differentiate its BPS equation. After this differentiation we need to evaluate the derivatives of $\beta$ and $\tilde\beta$. Taking into account (\ref{betas-alphas}), one gets:
\beq
\beta'\,=\,-\tilde\beta\alpha'\,\,,\qquad\qquad
\tilde\beta'\,=\,\beta\alpha'\,\,,
\eeq
and, after using the value of $\alpha'$ from (\ref{alpha-prime}), one obtains:
\bear
&&\beta'\,=\,\frac{3}{4}\,\,\Big[\,
N_c\gamma'e^{-g-h}+\frac{3}{4}w'e^{g-h}\,\Big]\,\tilde\beta\,\,,\rc\rc
&&\tilde\beta'\,=\,-\frac{3}{4}\,\,\Big[\,
N_c\gamma'e^{-g-h}+\frac{3}{4}w'e^{g-h}\,\Big]\,\beta\,\,.
\eear
Moreover, from our ansatz the term containing the RR three-form $F_3$ in the equation of motion of the dilaton is:
\bear
&&{e^{{3\phi\over 2}}\over 12}\,\,F_3^2\,=\,{3\over 8}\,N_c^2\,(\gamma')^2\,
e^{-2g-2h}\,+\,
2N_c^2\,e^{-6g}\,+\,{3\over 8}\,N_c^2\,\Big[\,1\,-{4N_f\over N_c}\,+\,w^2\,-\,2 w\gamma\,\Big]^2\,\,e^{-2g\,-\,4h}\,+\,\rc\rc
&&\qquad\qquad\qquad\qquad\qquad\qquad
+\,{3\over 2}\,N_c^2\,(w-\gamma)^2\,e^{-4g-2h}\,+\,{N_c^2\over 32}\,
 V^2\,e^{-6h}\,\,.
\eear
Amazingly, after a very long calculation one gets the simple result:
\beq
\phi''\,+\,2\Big(\phi'\,+\,3(h'\,+\,g')\Big)\phi'\,-\,{e^{{3\phi\over 2}}\over 12}\,\,F_3^2\,=\,
6N_f\,e^{-2h-2g}\,\,.
\eeq
In order to verify that the right-hand side of (\ref{dil-eom}) reproduces exactly this result, let us use the action  of the smeared flavor branes in terms of the charge distribution four-form $\Omega$. From the expression of this action it is straightforward to evaluate its contribution  to the equation of motion of the dilaton:
\beq
-\,\frac{2\kappa^2_{10}}{\sqrt{-G}} \,\,\,\,e^{{\phi\over 2}}\,\,
\frac{\delta}{\delta \phi} S_{flavor}\,=\,2\,\pi^2\,e^{\phi}\,
\sum_i\,\Big|\,\Omega^{(i)}\,\Big|\,=\,6N_f\,e^{-2h-2g}\,\,,
\eeq
which proves that, indeed, the equation of motion of the dilaton is satisfied.

Let us next verify that  the Einstein equations are satisfied. These equations are:
\begin{eqnarray}   
R_{MN}\,-\,\frac{1}{2} G_{MN}R\,&=&\, \frac{1}{2} \left ( \partial_M \phi \partial_N \phi \,-\,\frac{1}{2}G_{MN}\partial_P\phi \partial^P \phi \right ) \,+\, \,\, \rc
&+& \frac{1}{12}e^{\phi} \left (  3F^{(3)}_{MPQ}F^{(3)PQ}_N\,-\,\frac{1}{2}G_{MN}F^2_{(3)} \right )\,+\,T_{MN} \,\, ,
\label{Eeq}
\end{eqnarray}
where $T_{MN}$ is the DBI contribution to the energy-momentum tensor, namely:
\begin{equation}
T_{ M N}\,=\, \frac{2\kappa^2_{10}}{\sqrt{-G}} \frac{\delta S_{DBI}}{\delta G^{MN}} \,\, .
\end{equation}

Taking into account that the absolute value of the smearing form $\Omega$ is defined in
(\ref{modulusOmega}), we can simply compute the DBI contribution to the energy-momentum tensor, in flat components:
\begin{equation}
T_{\hat M \hat N}\,=\,-2\pi^2\,e^{\phi/2} \left [  \eta_{\hat M \hat N} \sum_{i}^3 \mid \Omega^{(i)} \mid \,-\, \sum_{i}^3 \frac{1}{3! \mid \Omega^{(i)} \mid } (\Omega^{(i)})_{ \hat M \hat P \hat Q \hat R} 
(\Omega^{(i)})_{\hat N\hat S \hat T \hat U}\eta^{\hat P \hat S}
\eta^{ \hat  Q\hat T}\eta^{\hat R \hat U}\right ] \,\, .
\end{equation}
In components, it means that
\begin{eqnarray}
&&T_{x_{\mu}x_{\nu}}\,=\,T_{rr}\,=\,-6N_f\,e^{-2f-2h-2g} \,\, , \rc\rc
&&T_{ii}\,=\, T_{\hat{i}\hat{i}}\,=\,-2N_f\,e^{-2f-2h-2g}\,\, ,
\end{eqnarray}
with  the off-diagonal components being zero. 

Moreover, from the value of the spin connection we can easily compute the  different components of the Ricci tensor. 
In  flat indices these components  are the following:
\begin{eqnarray}
R_{x^{\mu}x^{\nu}}\,&=&\,-\eta_{x^{\mu}x^{\nu}}e^{-2f} \left [ f''\,+\, 8(f')^2 \,+\, 3 f' h' \,+\, 3 f' g' \right ] \,\, , \rc
R_{rr}\,&=&\, -3e^{-2f} \left [  3 f'' \,+\, h'' \,+\, g''\,+\, h'(f'+h')\,+\,g'(f'+g')\,+\,\frac{1}{8}e^{2g-2h}(w')^2 \right ] \,\, , \rc 
R_{ii}\,&=&\,-e^{-2f} \Big [ f''\,+\,h''\,+\,8(f')^2\,+\,3(h')^2\,+\,11f' h'\,+\,3f' g' \,+\,3 h' g' \,+\,\, \rc
&+&\, e^{2g-4h} \frac{(1-w^2)^2}{4}\,+\, e^{2g-2h} \frac{(w')^2}{8}\,-\,2e^{-2h} \Big ] \,\, , \rc
R_{i\hat{i}}\,&=&\,e^{-2f+g-h} \left [  \frac{w''}{4}\,+\,2w' f'\,+\,\frac{5}{4} w' g' \,+\,\frac{1}{4} w' h' \,+\,e^{-2h}\frac{w(1-w^2)}{2}  \right ] \,\, , \rc
R_{\hat{i}\hat{i}}\,&=&\,-e^{-2f} \Big [ f''\,+\,g''\,+\,8(f')^2\,+\,3(g')^2\,+\,11f' g'\,+\,3f' h' \,+\,3 h' g' \,-\,\, \rc
&-&\, e^{2g-4h} \frac{(1-w^2)^2}{8}\,-\, e^{2g-2h} \frac{(w')^2}{8}\,-\,2e^{-2g} \Big ] \,\, ,
\end{eqnarray}
with the remaining components zero. It is straightforward to calculate now the curvature scalar in Einstein frame:
\begin{eqnarray}
R\,&=& \, -e^{-2f} \Big [ 18f''\,+\,6h''\,+\,6g''\,+\,72(f')^2\,+\,12(h')^2\,+\,12(g')^2\,+\,54f' h' \,+\,54f' g' \,+\,18 h' g' \,+\, \,\, \rc
&+& \frac{3}{8} e^{2g-2h}(w')^2\,+\, \frac{3}{8}e^{2g-4h}(1-w^2)^2\,-\,6e^{-2h}\,-\,6e^{-2g} \Big ]\,\, .
\end{eqnarray}
Finally, one can verify that the BPS equations we have found imply the fulfillment of eq. (\ref{Eeq}) along its different components. 

Finally, it remains to verify the equation of  motion for the RR three-form $F_3$, which is just:
\beq
d\,\big(\,e^{\phi}\,{}^*F_3\,\big)\,=\,0\,\,.
\label{d*F3}
\eeq
After computing the Hodge dual of $F_3$ with the metric (\ref{ansatz}), one can demonstrate that (\ref{d*F3}) is equivalent to the following second-order differential equation for the functions of our ansatz:
\bear
&&e^{h+g}\,\big[\,\gamma''\,+\,(2\phi'\,+\,h'\,+\,g'\,)\,\gamma'\,\big]\,+\,
4\,e^{h-g}\,\big(\,w-\gamma\,\big)\,+\,{e^{3(g-h)}\over 4}\,\,\big(\,1\,-\,w^2\,\big)\,V\,+\,
\rc\rc
&&\qquad\qquad\qquad+\,
2\,e^{g-h}\,\Big(\,1\,-\,{4N_f\over N_c}\,+\,w^2\,-\,2w\,\gamma\,\Big)\,w\,=\,0\,\,,
\eear
which, again, can be shown to be a consequence of our first-order BPS equations.

\medskip

%%%%%%%%%%%%%%%%%%%%%%%%%%%%%%%%%%%%%%%%%%%%%%%%%%%%%%%%%%%%%%%%%%%%%%%%%%

\end{document}